\documentstyle[12pt,epsf]{article}

\newcommand{\mcemptyl}{\multicolumn{2}{|c}{\mbox{}}}
\newcommand{\mcemptyr}{\multicolumn{2}{c|}{\mbox{}}}
\newcommand{\mcemptyll}{\multicolumn{2}{||c}{\mbox{}}}
\newcommand{\mcemptyrr}{\multicolumn{2}{c||}{\mbox{}}}

\newcommand{\plus}{\makebox[13pt][c]{$+$}}
\newcommand{\minus}{\makebox[13pt][c]{$-$}}
 
\newcommand{\err}[2]{\raisebox{0.01em}{\scriptsize
{$\,\begin{array}{@{}l@{}}
                          \plus\makebox[0.55em][r]{#1} \\[-0.12em]
                          \minus\makebox[0.55em][r]{#2}
                        \end{array}$}}}
\newcommand{\er}[2]{\raisebox{0.01em}{\scriptsize
{$\,\begin{array}{@{}l@{}}
                          \plus\makebox[0.15em][r]{#1} \\[-0.12em]
                          \minus\makebox[0.15em][r]{#2}
                        \end{array}$}}}
\newcommand{\tr}{\mbox{Tr}}

\newcommand{\geqsim}{\,\raisebox{-0.6ex}{$\buildrel > \over \sim$}\,}
\newcommand{\leqsim}{\,\raisebox{-0.6ex}{$\buildrel < \over \sim$}\,}
\newcommand{\gev}{{\rm GeV}}

\let\a=\alpha \let\be=\beta  
   
  \let\la=\lambda

\def\0{\over } 
\def\1{\vec }     
\def\2{{1\over2}} 
\def\4{{1\over4}}            
\def\5{\bar }  
\def\6{\partial } 
\def\7#1{{#1}\llap{/}}                         
\def\8#1{{\textstyle{#1}}}
\def\9#1{{\bf {#1}}}                           
 
\def\llp{\hbox to 0pt{\hss/\hskip1.5pt}}
\def\llo{\hbox to 0.2pt{\hss /}} \def\llq{\hbox to 0pt{\hss/\hskip0.5pt}}
\def\so{\supset\hbox to 0pt{\hss $\displaystyle -$\hskip1pt}}

\def\<{\langle } \def\>{\rangle }

\def\i{{\rm i}} 
\let\nn=\nonumber  
\def\bea{\begin{eqnarray}} \def\eea{\end{eqnarray}} 
\def\beann{\begin{eqnarray*}} \def\eeann{\end{eqnarray*}} 
\def\beq{\begin{equation}} \def\eeq{\end{equation}}  
\begin{document} 
\setlength{\baselineskip}{18pt}                                     
\thispagestyle{empty}
\begin{flushright}
{\tt OUTP-97-44P\\ HD-THEP-97-37\\September 1997}
\end{flushright}
\vspace{5mm}
\begin{center}
{\Large \bf
 Scalar-gauge dynamics in (2+1) dimensions at small and
 large scalar couplings}\\ \vspace{15mm}
\renewcommand{\thefootnote}{\fnsymbol{footnote}}
{\large O.~Philipsen$^{1}$, M.~Teper$^{2}$
 and H.~Wittig$^{2,}$\footnote{PPARC Advanced Fellow}}\\
\renewcommand{\thefootnote}{\arabic{footnote}}
 \vspace{10mm}
{\it 
$^{1}$ Theoretische Physik, Universit\"at Heidelberg \\ Philosophenweg 16, 
D-69120 Heidelberg, Germany\\
$^{2}$ Theoretical Physics, University of Oxford \\1 Keble Road,
            Oxford OX1 3NP, U.K.}

\end{center}
\vspace{2cm}

\begin{abstract}
\thispagestyle{empty}
\noindent
We present the results of a detailed calculation of the excitation
spectrum of states with quantum numbers $J^{PC}=0^{++}, 1^{--}$ and
$2^{++}$ in the three-dimensional SU(2) Higgs model at two values of
the scalar self-coupling and for fixed gauge coupling. We study the
properties of Polyakov loop operators, which serve to test the
confining properties of the model in the symmetric phase. At both
values of the scalar coupling we obtain masses of bound states
consisting entirely of gauge degrees of freedom (glueballs),
which are very close to those obtained in the pure gauge theory. We
conclude that the previously observed, approximate decoupling of the
scalar and gauge sectors of the theory persists at large scalar
couplings. We study the crossover region at large scalar coupling and
present a scenario how the confining properties of the model in the
symmetric phase are lost inside the crossover by means of flux tube
decay. We conclude that the underlying dynamics responsible for the
observed dense spectrum of states in the Higgs region at large
couplings must be different from that in the symmetric phase.
\end{abstract} 
\setcounter{page}{0}

\newpage

\section{Introduction}

Over the past few years the SU(2) Higgs model in three dimensions has
been the subject of many thorough numerical studies by means of
lattice Monte Carlo simulations \cite{kaj95} -- \cite{Bielefeld}. The
main incentive for these investigations was to clarify the nature of
the electroweak phase transition in the framework of the dimensional
reduction programme \cite{fkrs94}, where it is much easier to obtain
precise numerical results than in the full four-dimensional theory at
finite temperature \cite{mont97}. In these studies it has been
established that the electroweak phase transition is weakly first
order for Higgs masses up to $m_H\sim 70\,\gev$, while for Higgs
masses $m_H \geqsim 80\,\gev$ the transition disappears and turns into
a smooth crossover \cite{bp94,kaj96}. Hence, the Higgs region and the
confinement region in parameter space are indeed analytically
connected, as was conjectured a long time ago for models with
fixed-length Higgs fields \cite{frad79}. A review of the existing
lattice results as well as a comparison with perturbation theory can
be found in \cite{rum96}.

While the nature of the phase diagram and the order of the phase
transition are now well determined, our understanding of the
symmetric phase is still incomplete. Calculations of the mass spectrum
based on the use of gauge-invariant operators lead to a picture of a
confining symmetric phase with a dense spectrum of bound states. 
Recent attempts to describe the spectrum in terms of bound state models
may be found in \cite{do95,bp97}.
This part of the phase diagram is of great intrinsic interest because it
exhibits the same qualitative behaviour as QCD. 
In this region the non-Abelian gauge theory has confining properties,
yet the presence of matter fields in the fundamental representation
leads eventually to the screening of charges at large distances.
This apparent similarity between QCD and the symmetric phase of
scalar-gauge models opens the possibility to study some aspects of QCD
in a model that is computationally much less demanding.

In the Higgs phase, on the other hand, for small enough scalar
couplings a perturbative calculation of the spectrum is applicable and
leads to results in good agreement with those from lattice
simulations. Due to the analytic connectedness of the phase diagram
all mass eigenstates in the Higgs region may be continuously mapped
into their counterparts in the confinement region. This offers the
possibility to study the onset of non-perturbative physics and
confinement by smoothly moving from the Higgs region into the
confinement region of the phase diagram.

In this paper we elaborate on our previous computation of the mass
spectrum of the SU(2) Higgs model in 2+1 dimensions\,\cite{us}. Here,
however, we move away from the original context of the electroweak
phase transition, and our main interest lies in the dynamics of the
three-dimensional theory and what it might have in common with QCD. In
particular, we want to obtain a more complete picture of the
excitation spectrum and investigate in more detail the previously
observed approximate decoupling of the scalar and gauge sectors of the
model\,\cite{us}. Thus, in addition to the $0^{++}$ and $1^{--}$
states considered before, we now also compute correlations of $2^{++}$
and Polyakov loop operators. Although our analysis does not yet
incorporate multiparticle states, it leads to a more complete picture
of the symmetric phase.

In contrast to our previous work, where we carefully studied the
approach to the continuum limit, we now consider the theory at a fixed
value of the gauge coupling, i.e. $\beta_G=9$, where our earlier
results show that the physics is already very close to the continuum
limit. Furthermore, at large scalar coupling it has been demonstrated
numerically\,\cite{kaj96} that the disappearance of the phase
transition survives in the continuum limit and that no significant
lattice artefacts are to be expected for $\beta_G=9$.

Our main results are as follows. The previously observed approximate
decoupling of states composed purely out of gauge degrees of freedom
in the $0^{++}$ channel\,\cite{us} extends to the $2^{++}$
channel. The masses of these states are very close to their values in
the pure gauge theory. Furthermore, the decoupling occurs at both
small and large values of the scalar self-coupling, and in the latter
case it can be observed also deep in the crossover region. Second, at
large scalar coupling we find that a dense spectrum of states is
observed even in the Higgs phase, yet the underlying dynamics must be
quite different, as confinement is no longer observed. We present a
scenario how the confining properties of the theory are lost inside
the crossover region: within the QCD context it is natural to use the
concept of flux tubes in the symmetric phase as a manifestation of
confinement. As one enters the crossover region the flux tube develops
a rapidly growing decay width and eventually becomes so unstable that
there is no longer any remnant of linear confinement (i.e. a range of
distances where the static potential shows a linear behaviour).
Finally, the inclusion of correlations of Polyakov loops changes some
of our previous conclusions about the influence of finite-volume
effects. In particular, we find that finite-size effects, especially
in the lowest $0^{++}$ state are less severe than was claimed
before\,\cite{us}.

Some of the results presented in this paper have been summarised
elsewhere\,\cite{lat96,owe_eger_97}. We also draw the reader's
attention to some recent related work\,\cite{GuIlSchiStr97}.

The remainder of this paper is organised as follows. In section\,2 we
briefly summarise the lattice techniques used in this work. Section\,3
contains a discussion and comparison of the properties of the Higgs
and confinement phases for small scalar self-coupling. In section\,4
we describe how the properties of the Higgs and confinement regions of
the phase diagram change when the size of the scalar self-coupling is
increased. In section\,5 we follow the analytic connection of the mass
spectrum through the crossover. Results indicating flux tube decay in
the crossover region are presented in section\,6, and section\,7
contains our conclusions.

\section{Methodology}

We work with the three-dimensional lattice action given by
\bea \label{actlat}
S[U,\phi]&=&\be_G\sum_p\left(1-\frac{1}{2}\tr U_p\right)
+\sum_x\Bigg\{-\be_H\sum_{\mu=1}^3\frac{1}{2}
\tr\Big(\phi^{\dagger}(x)U_\mu(x)
\phi(x+\hat{\mu})\Big)  \nn\\
&&+\frac{1}{2}\tr\Big(\phi^{\dagger}(x)\phi(x)\Big)
+\be_R\left[\frac{1}{2}\tr\Big(\phi^{\dagger}(x)\phi(x)\Big)-1\right]^2
\Bigg\},
\eea
where $U_\mu(x)\in\rm SU(2)$ is the link variable, $U_p$ denotes the
plaquette, and $\phi(x)$ is the scalar field. The bare parameters
$\beta_G,\,\beta_H$ and $\beta_R$ are the inverse gauge coupling, the
scalar hopping parameter, and scalar self-coupling, respectively. In
our simulations we fix the combination of parameters
\beq
 \frac{\lambda_3}{g_3^2} = \frac{\beta_R\,\beta_G}{\beta_H^2} 
\;,
\eeq
where $\lambda_3/g_3^2$ is the ratio of the scalar self-coupling and
the gauge coupling in the continuum formulation.  Exact relations
between the lattice and the continuum parameters at the two-loop level
governing the approach to the continuum limit can be found in
\cite{lai95}. In this work we consider the cases of small,
i.e. $\lambda_3\ll g_3^2$ and large, i.e. $\lambda_3\sim O(g_3^2)$
scalar self-coupling, by choosing $\lambda_3/g_3^2=0.0239$ and
$\lambda_3/g_3^2=0.2743$, respectively. At the small value, which was
used in our earlier study\,\cite{us}, the confinement and Higgs phases
are separated by a first order phase transition, whereas for the
larger value of $\lambda_3/g_3^2$ the transition has turned into a
smooth crossover\footnote{Note that at tree level, the ratio
$\lambda_3/g_3^2$ is proportional to the ratio of Higgs and $W$-boson
masses in the four-dimensional theory at finite temperature. As an
orientation for those readers who are familiar with studies of the
electroweak phase transition, we add that in the framework of
dimensional reduction, our values of $\lambda_3/g_3^2$ correspond to
tree-level, zero temperature Higgs masses of $35\,\gev$ and
$120\,\gev$, respectively\,\cite{kaj95}.}.

\subsection{The blocking procedure}

In any mass calculation on the lattice it is important to have a good
projection property of the interpolating operators in order to obtain
a reliable signal. For gauge theories with or without scalar fields it
has been demonstrated that the projection of the basic operators onto
the desired state can be considerably enhanced by employing so-called
``blocking'' or ``fuzzing''
techniques\,\cite{tep87,albanese,how89,us}.

As in our previous work\,\cite{us}, we construct blocked link
variables $U_\mu^{(n)}(x)$ at blocking level~$n$ according to the
procedure originally defined in\,\cite{tep87}. Furthermore, a blocked
version of the scalar field, $\phi^{(n)}(x)$ at blocking level~$n$ is
constructed according to
\beq
\phi^{(n)}(x)=\frac{1}{5}\left\{\phi^{(n-1)}(x)+\sum_{j=1}^{2}\left[
U_j^{(n-1)}(x)\phi^{(n-1)}(x+\hat{\j})+U_j^{(n-1)\dag}(x-\hat{\j})
\phi^{(n-1)}(x-\hat{\j})\right]\right\}\;.
\eeq
Note that we will take correlations in the $\hat{3}$ direction and so
our blocking always remains within the $(1,2)$-plane. In addition, we
consider blocked scalar fields $\phi^{(n,j)}(x)$ with non-local
contributions from one spatial direction~$j$ only:
\bea
\phi^{(n,j)}(x)=\frac{1}{3}\Big\{\phi^{(n-1,j)}(x) & + & 
U_j^{(n-1)}(x)\phi^{(n-1,j)}(x+\hat{\j})  \nonumber\\
& + &U_j^{(n-1)\dag}(x-\hat{\j})\phi^{(n-1,j)}(x-\hat{\j})\Big\},\quad
j=1,2.
\label{EQasymblock}
\eea
It is obvious that all blocking procedures employed are gauge
invariant and can be iterated in order to create a large set of
operators of various spatial extensions in an efficient manner. In
practice, the blocking levels have to be tuned in order to identify
the operators with the best projection properties.

\subsection{Basic operators}
\label{SECop}
We now proceed to list the basic operators used in our simulation.
Since we use a larger set of operators compared to our previous
calculation, we have decided to change our notation slightly for
clarity. Although in general we use blocked links and scalar fields,
we suppress the superscripts which label the blocking level in this
subsection, except when stated explicitly.

After decomposing the scalar field $\phi(x)$ as
\beq
  \phi(x)=\rho(x)\alpha(x),\quad \rho(x)\ge0,\quad\alpha(x)\in\rm
  SU(2),
\eeq
we consider the following basic operators involving only scalar fields
or combinations of scalar and gauge degrees of freedom:
\bea
R(x) & \equiv &\frac{1}{2}\tr\left(\phi^{\dagger}(x)\phi(x)\right),
 \nonumber \\
L_j(x) & \equiv & 
\frac{1}{2}\tr\left(\a^{\dag}(x)U_j(x)\a(x+\hat{\j})\right),\quad j=1,2
 \nonumber \\
V_j^a(x) & \equiv & 
\frac{1}{2}\tr\left(\tau^a\a^{\dag}(x)U_j(x)\a(x+\hat{\j})\right),\quad
j=1,2,
\label{EQ_rlv}
\eea
where $\tau^a$ is a Pauli matrix. Note that at this stage we have not
yet specified the quantum numbers $J^{PC}$ of the operators. This is
postponed to the end of this subsection when all basic operators have
already been defined.

We also consider Wilson loops $C_{ij}$, i.e. gauge invariant operators
constructed out of link variables only, e.g.
\beq
 C_{ij}^{1\times1}(x)=\tr\left(
 U_i(x)U_j(x+\hat{\i})U^\dag_i(x+\hat{\j})U^\dag_j(x) \right),
 \quad i,j=1,2,\,i\not=j,
\label{EQ_c}
\eeq
and in addition to the elementary plaquette $C^{1\times1}$, we also
consider squares of size $2\times2$ as well as rectangles of size
$1\times2$, $1\times3$, $2\times3$. Hence, we have five versions of
Wilson loop operators, viz.
\beq
  C_{ij}^{1\times1},\;C_{ij}^{2\times2},\;C_{ij}^{1\times2},\;
  C_{ij}^{1\times3},\;C_{ij}^{2\times3}.
\eeq
As we shall see, a larger set of this type of operator will
significantly improve our ability to extract information about the
gauge sector of the theory.

An operator which is useful to probe the confining properties of a
theory is the Polyakov loop, i.e. a loop of length $L$ of link
variables that winds around the spatial dimensions of the lattice,
viz.
\beq
  P_j^{(L)}(x)=\tr\prod_{m=0}^{L-1}\,U_j(x+m\hat{\j}), \quad j=1,2.
\label{EQ_polya}
\eeq
The expectation value of the Polyakov loop vanishes if the theory is
confining, and it projects onto the state consisting of a
chromoelectric flux loop that closes through the boundary in the
$\hat\j$ direction. Correlations of $P_j^{(L)}$ can then be used to
extract a string tension from the lowest mode of its exponential
fall-off, the so-called ``torelon'' mass. However, in a theory with
matter fields, such as ours, one eventually expects the string to
break beyond some large separation, due to pair creation (just as in
QCD). In our model we also have a phase with spontaneously broken
symmetry, the Higgs phase. Here, this interpretation of the Polyakov
loop correlator no longer holds, the Polyakov loop has a large
expectation value, and the torelon mass cannot be used to extract a
string tension. Instead, since a weak coupling expansion of the
Polyakov loop correlator applies in the Higgs phase, its leading term
corresponds to an exchange of two spin-1 particles (i.e. $W$-bosons),
and therefore the lightest mass above the vacuum should be interpreted
as a two-$W$ scattering state.

Apart from being relevant for the study of confinement, the Polyakov
loop also plays an important r\^ole in understanding apparent
finite-size effects in glueball calculations in the pure gauge
theory\,\cite{cm87a}. Since we are interested in the detailed
structure of the spectrum, and, in particular, higher excitations, we
need to address the influence of torelons in the various spin
channels. Although single torelons, arising from correlations of the
Polyakov loop $P_j^{(L)}$ cannot contribute directly to correlations
of operators $C^{1\times1},\ldots, C^{2\times3}$, they may well be
relevant for the operators $R(x)$ and $L_j(x)$. This is because of the
iterative, non-local nature of the blocking procedure, such that
highly blocked versions of $R(x)$ and $L_j(x)$ may contain
contributions which, for periodic boundary conditions, wind around the
lattice and may thus have a sizeable projection onto
$P_j^{(L)}$. Furthermore, torelon-antitorelon pairs are known to give
rise to large finite-volume effects in glueball
calculations\,\cite{cm87a}. Apart from studying correlations of
Polyakov loops {\it per se\/}, we have decided to construct
single-torelon and torelon-pair operators in the $0^{++}$ and $2^{++}$
channels and to study their correlations as a safeguard against
finite-size effects.

We now give a complete list of the types of operators with quantum
numbers $J^{PC}$ and also introduce the symbols which will later label
the contributions from the various operators to a given state.

\begin{tabbing}
\indent \= $P_d$:\,\, \= \kill
$0^{++}$ channel: \\
\> $R$:   \> $R$ \\
\> $L$:   \> $L_1+L_2$ \\
\> $C$:   \> symmetric combinations of $C^{1\times1},\,C^{2\times2},\,
             C^{1\times2},\,C^{1\times3},\,C^{2\times3}$\\
\> $P$:   \> $P_1^{(L)}+P_2^{(L)}$ \\
\> $P_d$: \> $P_1^{(L)}\cdot P_2^{(L)}$ \\
\> $T$:   \> $\left(P_1^{(L)}\right)^2+\left(P_2^{(L)}\right)^2$ \\
\> \\
$1^{--}$ channel: \\
\> $V$:   \> $V_j^a,\quad j=1,2\quad a=1,2,3$ \\
\> \\
$2^{++}$ channel: \\
\> $R$:   \>
$R^{(n)}_{-}\equiv\textstyle\frac{1}{2}
             \left\{\tr\big[\phi^{(n,1)\dag}
                            \phi^{(n,1)}\big]
                   -\tr\big[\phi^{(n,2)\dag}
                            \phi^{(n,2)}\big] \right\},\quad n>1$ \\
\> $L$:   \> $L_1-L_2$ \\
\> $C$:   \> antisymmetric combinations of 
             $C^{1\times2},\,C^{1\times3},\,C^{2\times3}$\\
\> $P$:   \> $P_1^{(L)}-P_2^{(L)}$ \\
\> $T$:   \> $\left(P_1^{(L)}\right)^2-\left(P_2^{(L)}\right)^2$ \\
\end{tabbing}
%
Here, the operators are understood to be zero-momentum sums of the
corresponding operators defined in
eqs.\,(\ref{EQ_rlv})--(\ref{EQ_polya}).
The labels $R,\,L$ refer to operators containing scalar fields,
whereas the Wilson loop operators~$C$ are composed only out of gauge
degrees of freedom. The labels $P,\,P_d$ denote torelon operators, and
$T$ labels pairs of torelons in the respective channels.  Note that
the definition of the operator $R^{(n)}_{-}$ in the $2^{++}$ channel
is only possible with the kind of blocking defined in
eq.\,(\ref{EQasymblock}).

Our list of operators does not contain multiparticle operators, and
therefore we do not have control over continuum states in the
spectrum\footnote{Note that on a finite volume, those states have a
discrete spectrum.}. Here we wish to remark that we have evidence that
continuum states have only small overlaps on the basis of operators
used in this work, so that we are still able to extract the bound
state mass spectrum reliably.

\subsection{Matrix correlators}

In order to compute the excitation spectrum of states with given
quantum numbers, we construct matrix correlators by measuring all
cross correlations between different types of operators at several
blocking levels\,\cite{us}. This correlation matrix can then be
diagonalised numerically following a variational method. For a given
set of $N$ operators ${\phi_i}$ we find the linear combination that
minimises the energy, corresponding to the lightest state.

The first excitation can be found by applying the same procedure to
the subspace $\{\phi_i\}^{'}$ which is orthogonal to the ground 
state. This may be continued to higher states so that we end up with a set
of~$N$ eigenstates $\Phi_i,\,i=1,\ldots,N$ given by
\beq \label{eigen}
\Phi_i =  \sum_{k=1}^N a_{ik}\phi_k.
\eeq
The coefficients $a_{ik}$ quantify the overlap of each individual
operator $\phi_k$ used in the simulation onto a particular approximate
mass eigenstate $\Phi_i$. For a complete basis of operators this
procedure is exact. In practice, the quality of the approximation
clearly depends on the number~$N$ of original operators and their
projection properties. The determined eigenstates are removed from the
basis for the higher excitations, so that the basis for the latter
gets smaller and the corresponding higher states are determined less
reliably.

For our main spectrum calculation we work with $N=24$ operators in the
$0^{++}$ channel, i.e. $\phi_k\in R, L, C, P, P_d, T$, each considered
for at least two blocking levels. In the $2^{++}$ channel we choose
$N=23$ and $\phi_k\in R, L, C, P, T$. In the symmetric phase this
typically enables us to obtain mass estimates for the first 10--12
states in each of these channels.

For spin-1 particles, we construct a $4\times4$ matrix correlator from
the operator of type~$V$ considered at four blocking levels. Thereby
we can extract masses for the first two or three states only. In the
computation of separate correlations of the Polyakov loop operator
itself, we also compute a $4\times4$ matrix correlator using
different blocking levels.

For a more detailed discussion of the variational method based on
matrix correlators, along with numerical investigations to test its
efficiency, we refer to our previous work\,\cite{us}.

\subsection{Simulation and analysis details}
\label{SECanalysis}

The Monte Carlo simulation of the lattice action in
eq.\,(\ref{actlat}) is performed using the same algorithms as in our
previous work. The bare parameters $\beta_G,\,\beta_H$ and $\beta_R$
are fixed by applying the constraints $\lambda_3/g_3^2=0.0239$ and
$\lambda_3/g_3^2=0.2743$, respectively, and using the same procedure
as in section\,3 of ref.\,\cite{us}. For the update of gauge variables
we use a combination of the standard heatbath and over-relaxation
algorithms for SU(2)\,\cite{fabhaan,kenpen}. The scalar degrees of
freedom are updated using the algorithm described
in\,\cite{bunk_lat94}. In general, the applicability of this algorithm
is restricted to cases where the self-coupling $\beta_R$ is not too
large, as otherwise the acceptance rate decreases substantially. In
our simulation, however, we did not observe any significant drop in
the acceptance rate, even at our larger value of $\beta_R$. Thus the
algorithm may be applied safely in the region of the parameter space
explored in this paper.

As before, we define a ``compound'' sweep to consist of a combination
of one heatbath and several over-relaxation updates of the gauge and
scalar fields. In the Higgs and confinement regions of the phase
diagram at both small and large scalar coupling we typically gathered
about 12000 compound sweeps for our main spectrum calculation. In the
crossover region we also monitored the spectrum at several, closely
spaced values of $\beta_H$, for which about 5000 compound sweeps were
accumulated each. Thus, the statistics in our present study is much
lower than in our previous work. At first sight this may seem
surprising, since we now seek to obtain a more detailed structure of
the mass spectrum. However, we wish to remark that our basis of
operators is much larger than in\,\cite{us}, so that a loss in
statistics is compensated for by using the information from a greater
number of correlators in a particular channel. In addition, the
diagonalisation procedure requires a large basis of operators in order
to yield reliable results for the excitation spectrum.

Our standard lattice sizes in this study are $36^3$ and $24^3$, which,
on the basis of our previous study, do not lead to significant
finite-size effects in the $0^{++}$ and $1^{--}$ channels. In order to
study finite-volume effects in more detail, especially in the $2^{++}$
channel, we have also simulated lattices of size $16^2\cdot24$.

Our mass estimates are extracted from two-parameter, correlated,
single-exponential fits over a finite interval $[t_1,\,t_2]$ to
correlation functions of eigenstates $\Phi_i$ (c.f. subsection\,3.4 in
ref.\,\cite{us}). We have checked explicitly that the alternative
fitting formula motivated in\,\cite{MontWeisz87}, where an additional
constant is considered, does not produce an appreciable difference in
our mass estimates.

In cases where the effective masses calculated from the correlation
function did not show a plateau long enough to perform a correlated
fit, we used effective masses for our best estimates.  This was also
our preferred method for the scan of the crossover region, where we
were more interested in qualitative features rather than very precise
mass estimates.  Results which were obtained from effective masses, or
from eigenstates whose individual overlaps $a_{ik}$ were small
throughout, are marked by an asterisk in our data tables in
sections\,\ref{SECsmall} and\,\ref{SEClarge}.

Statistical errors are estimated using a jackknife procedure for which
the individual measurements have been accumulated in bins of 500
sweeps. For some of our most precise results we also quote a
systematic error, which was obtained by quoting the difference between
our best estimate and the result from an uncorrelated fit, and, in
some cases, the result obtained using an alternative fitting interval.

\section{Results at small scalar coupling}
\label{SECsmall}
To explore small scalar coupling physics we fix $\la_3/g^2_3=0.0239$
as in ref.\,\cite{us}. For this value the system exhibits a strong
first order transition upon variation of $\be_H$ and the Higgs and
confinement properties on each side of the transition are very
pronounced and easily distinguishable. In the symmetric phase we
choose the point $\be_H=0.3438$ and in the Higgs phase $\be_H=0.3450$.
The $0^{++}$ and $1^{--}$ channels of the mass spectrum at these
parameter values have already been determined in our previous work
\cite{us}. Here we repeat this calculation with a larger basis of
operators in order to obtain more details about the excitation
spectrum. We do so at $\beta_G=9$, where we know from our previous
study that we are already close to the continuum limit of the theory.

\subsection{Mass eigenstates in the Higgs and confinement regions}

\begin{figure}[tbp]
\begin{center}
\leavevmode
\epsfysize=250pt
\epsfbox[20 30 620 730]{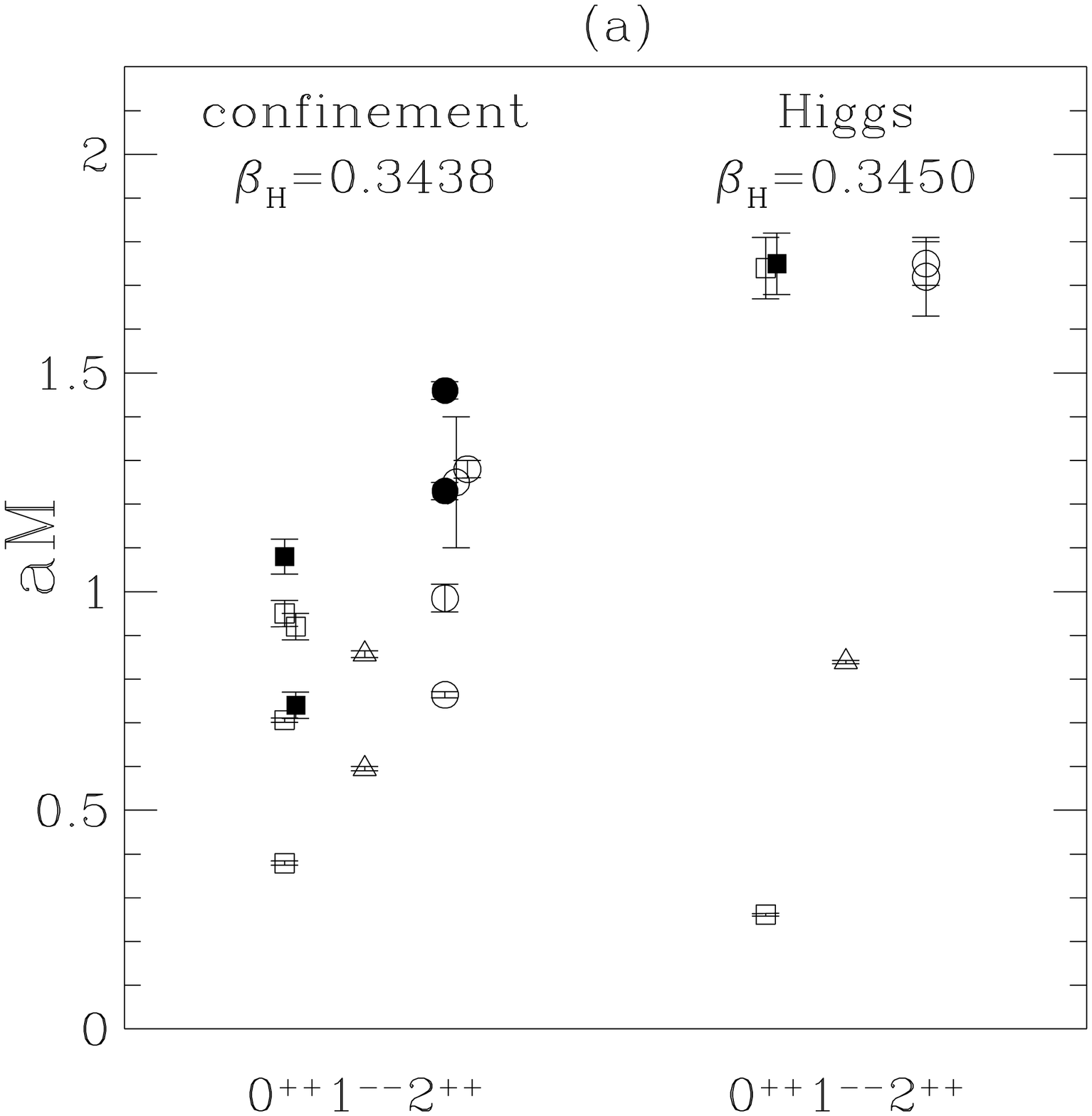}
\leavevmode
\epsfysize=250pt
\epsfbox[20 30 620 730]{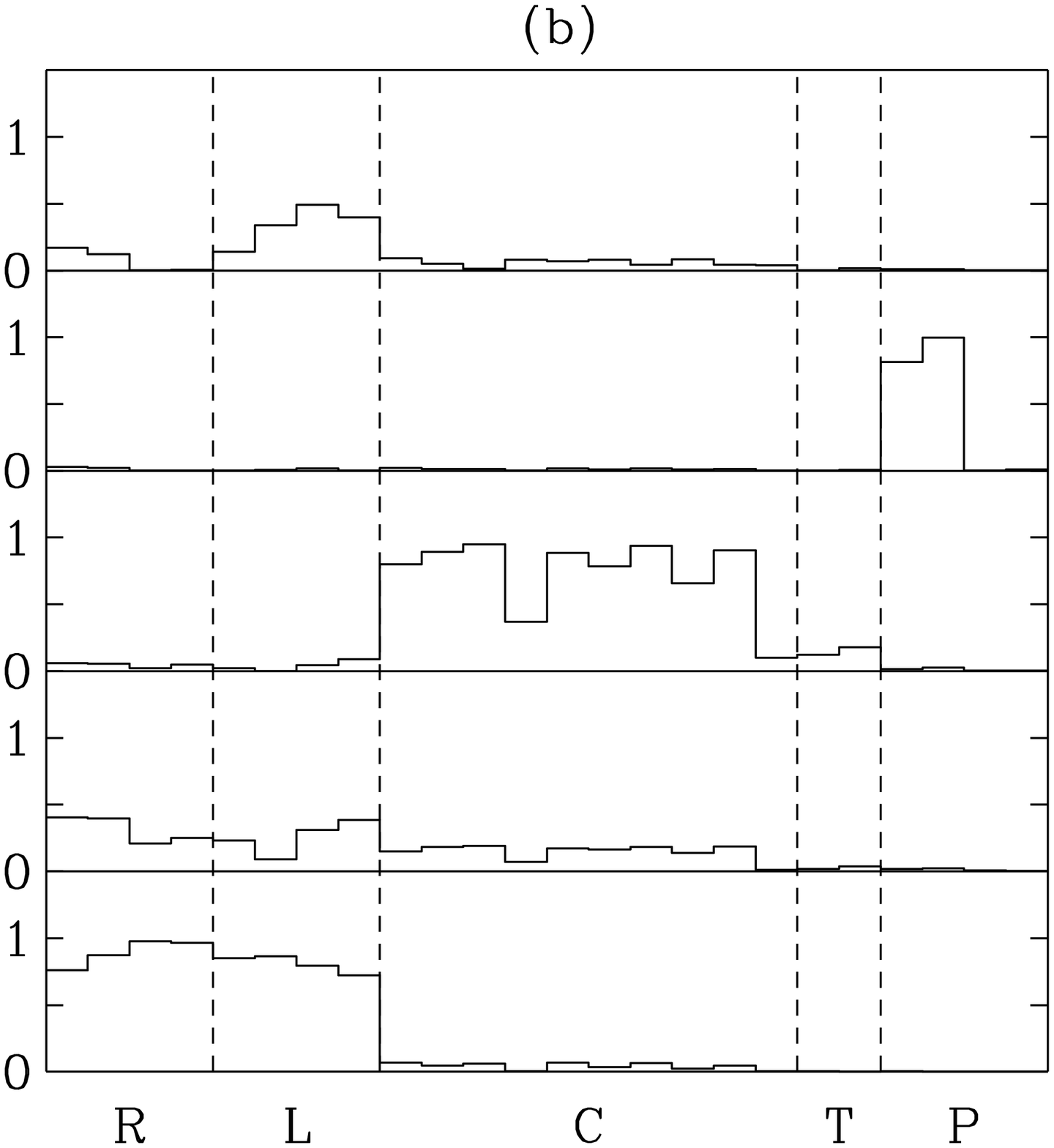}
\end{center}
\vspace{-1.6cm}
\caption[]{\label{spec_lsmall} {\it (a) The lowest states of the
spectrum in the confinement (left) and Higgs (right) region at small
scalar coupling, $\be_G=9$. The lattice sizes are $36^3$ in the
confinement and $24^3$ in the Higgs phase, respectively. Full symbols
indicate states that receive predominantly contributions from Wilson
loops. (b) The coefficients $a_{ik}$ of the 24 operators used in the
simulation on $36^3$ for the five lowest $0^{++}$ eigenstates in the
confinement phase.}}
\end{figure}

The lowest states of the mass spectrum at the two points investigated
are shown in Fig.\,\ref{spec_lsmall}\,(a). In the confinement region
one observes the now familiar, dense spectrum of states for the three
spin channels. As an illustration of the contributions of different
types of basic operators to the mass eigenstates, we plot the
coefficients $a_{ik}$ (c.f. eq.\,(\ref{eigen})) in
Fig.\,\ref{spec_lsmall}\,(b) for the lowest five eigenstates in the
$0^{++}$ channel in the confining phase.

Our main results in the Higgs phase are listed in
Table\,\ref{mhiggs_low}. Here, the only low-lying states are the
familiar Higgs and $W$-boson states, which are also well described by
perturbation theory. Note in particular that the lowest mass in the
$2^{++}$ channel is rather large and consistent with that of a
$W\,W$-scattering state -- as are the excited states in the $0^{++}$
channel. In particular, consider one of the first excited states in
the $0^{++}$ channel. This state has almost exclusively contributions
from Wilson loops, i.e. it consists mainly of gauge degrees of
freedom. This suggests interpreting it as a $W\,W$-scattering state,
and Table\,\ref{mhiggs_low} shows that the mass of this state is
indeed compatible with twice the $W$-mass. This result is also
expected from perturbation theory, where the term with the slowest
exponential fall-off corresponds to a two-$W$ exchange diagram. It is
natural to ask at this point why we do not see a two-Higgs state as
the first excitation, since such a state is lighter than the two-$W$
state. The reason is that our operators appear to have practically no
projection on such a state. In order to see a two-Higgs state it would
be necessary to study the correlations of an operator which by
construction couples to such a state in leading order, for instance
$\left(R(\vec{p}=0)-\langle R\rangle\right)^2$. This, however, is
beyond the scope of this work.

\begin{table}[tbp]
\begin{center}
\begin{tabular}{|c|r@{.}l|r@{.}l|c|}
\hline
\hline
 & \multicolumn{2}{c|}{$16^2\cdot24$} & \multicolumn{2}{c|}{$24^3$}
 & \\
\cline{2-5}
No. & \multicolumn{2}{c|}{$aM$} & \multicolumn{2}{c|}{$aM$}
 & $J^{PC}$ \\
\hline
\hline
$\Phi_1$ & 0&2666(17)\err{0}{27} & 0&2613(28) & $0^{++}$ \\
$\Phi_2$ & 1&65(6)               & 1&74(7)${}^*$ & $0^{++}$ \\
$\Phi_3$ & 1&81(8)               & 1&75(7)${}^*$ & $0^{++}$ \\
\hline
$\Phi_1$ & 0&837(3)\er{0}{4}     & 0&839(4) & $1^{--}$  \\
\hline
$\Phi_1$ & \multicolumn{2}{c|}{\mbox{}} & 1&75(5)${}^*$
         & $2^{++}$ \\
$\Phi_2$ & \multicolumn{2}{c|}{\mbox{}} & 1&72(9)${}^*$ & $2^{++}$ \\
\hline
\hline
\end{tabular}
\end{center}
\caption{\em \label{mhiggs_low}
Results for all three spin channels in the Higgs region at
$\be_H=0.3450$. The second error in some results is an estimate of
systematic errors as described in subsect.\,\ref{SECanalysis}.} 
\end{table}

Our main results in the symmetric phase are listed in
Tables\,\ref{m0_lsmall}--\ref{m1_lsmall}.
Here and in the following tables, the ordering of states $\Phi_i$ was
obtained during the diagonalisation procedure from the effective mass
of the corresponding correlator on the first timeslice.
With our larger basis of operators we are able to identify a large
number of states (e.g. up to 16 states in the $0^{++}$ channel), and
quite a number of them receive their dominant contributions from
Polyakov-loop or torelon-pair operators. Note in particular the change
in the operator content in the ground state of the $2^{++}$ channel as
the lattice size is increased. This will be discussed in more detail
in connection with finite-size effects in the next subsection. Our
results for the lowest masses are in agreement with our earlier study
(c.f. Tables\,2 and\,3 in\,\cite{us}), although we now obtain a
slightly lower value for the mass in the $1^{--}$ channel.

\begin{table}
\begin{center}
\begin{tabular}{|c|r@{.}lc|c|r@{.}lc|}
\hline
\hline
\multicolumn{8}{|c|}{$L^2\cdot T=36^3,\qquad \beta_H=0.3438$}  \\
\hline
No.  &  \multicolumn{2}{c}{$aM[0^{++}]$} & Ops.   &
No.  &  \multicolumn{2}{c}{$aM[0^{++}]$} & Ops.   \\
\hline
$\Phi_1$  & 0&379(5)\er{0}{7} & $R,L$  &  $\Phi_7$ 
 & 1&08(4)\er{2}{0} & $C$    \\
$\Phi_2$  & 0&706(5)\er{0}{5} & $R,L$  &  $\Phi_8$ 
 & 1&35(2)          & $C$    \\
$\Phi_3$  & 0&74(3)$^*$       & $C$    &  $\Phi_9$ 
 & 1&27(8)          & $P$   \\
$\Phi_4$  & 0&84(2)           & $P$    & $\Phi_{10}$ 
 & 1&47(3)\er{0}{2} & $P$   \\
$\Phi_5$  & 0&95(3)           & $L$    & $\Phi_{11}$ 
 & 1&50(4)          & $C$    \\
$\Phi_6$  & 0&92(3)\er{2}{0}  & $R$    & $\Phi_{12}$ 
 & 1&56(4)          & $C$    \\
\hline
\hline
\multicolumn{8}{|c|}{$L^2\cdot T=24^3,\qquad \beta_H=0.3438$}  \\
\hline
No.  &  \multicolumn{2}{c}{$aM[0^{++}]$} & Ops.   &
No.  &  \multicolumn{2}{c}{$aM[0^{++}]$} & Ops.   \\
\hline
$\Phi_1$  & 0&382(3)          & $R$    &  $\Phi_9$ 
 & 1&10(2)          & $P$   \\
$\Phi_2$  & 0&577(8)          & $P$    & $\Phi_{10}$ 
 & 1&27(2)$^*$      & $C,T$  \\
$\Phi_3$  & 0&690(14)         & $R$    & $\Phi_{11}$ 
 & 1&25(3)$^*$      & $L$    \\
$\Phi_4$  & 0&756(7)\er{6}{0} & $C$    & $\Phi_{12}$ 
 & 1&44(3)\er{0}{2} & $T$    \\
$\Phi_5$  & 1&02(2)           & $C$    & $\Phi_{13}$ 
 & 1&51(4)          & $C$    \\
$\Phi_6$  & 0&95(6)$^*$       & $P_d$  & $\Phi_{14}$ 
 & 1&33(3)          & $L$    \\
$\Phi_7$  & 1&03(2)           & $L,C$  & $\Phi_{15}$ 
 & 1&50(4)          & $T$    \\
$\Phi_8$  & 0&99(2)           & $P_d$  & $\Phi_{16}$ 
 & 1&55(6)          & $P_d$  \\
\hline
\hline
\multicolumn{8}{|c|}{$L^2\cdot T=16^2\cdot24,\qquad \beta_H=0.3438$}  \\
\hline
No.  &  \multicolumn{2}{c}{$aM[0^{++}]$} & Ops.   &
No.  &  \multicolumn{2}{c}{$aM[0^{++}]$} & Ops.   \\
\hline
$\Phi_1$  & 0&315(9)\err{13}{0}  & $P$   &  $\Phi_8$ 
 & 0&87(5)       & $L$  \\
$\Phi_2$  & 0&40(2)              & $S$   &  $\Phi_9$ 
 & 1&08(8)$^*$   & $C$  \\
$\Phi_3$  & 0&64(2)              & $P_d$ & $\Phi_{10}$ 
 & 1&45(3)$^*$   & $C$  \\
$\Phi_4$  & 0&723(8)\er{0}{8}    & $C$   & $\Phi_{11}$ 
 & 1&20(15)$^*$  & $P_d$ \\
$\Phi_5$  & 0&78(7)$^*$          & $S$   & $\Phi_{12}$ 
 & 1&49(3)       & $C$  \\
$\Phi_6$  & 0&96(5)              & $T$   & $\Phi_{13}$ 
 & 1&28(9)       & $L$  \\
$\Phi_7$  & 0&96(4)              & $P$   & & \multicolumn{2}{c}{\mbox{}} & \\
\hline
\hline
\end{tabular}
\caption{ \label{m0_lsmall}
{\em Mass spectrum and dominant contributions from the
operator basis in the $0^{++}$ channel at small scalar self-coupling
in the symmetric phase.}}
\end{center}
\end{table}
\begin{table}
\begin{center}
\begin{tabular}{|c|r@{.}lc|c|r@{.}lc|}
\hline
\hline
\multicolumn{8}{|c|}{$L^2\cdot T=36^3,\qquad \beta_H=0.3438$}  \\
\hline
No.  &  \multicolumn{2}{c}{$aM[2^{++}]$} & Ops.   &
No.  &  \multicolumn{2}{c}{$aM[2^{++}]$} & Ops.   \\
\hline
$\Phi_1$  & 0&764(7)            & $L$   &  $\Phi_6$
 & 1&46(2)        & $C$    \\
$\Phi_2$  & 0&888(8)\er{5}{0}   & $P$   &  $\Phi_7$
 & 1&25(15)$^*$   & $P$   \\
$\Phi_3$  & 0&985(32)           & $R,L$ &  $\Phi_8$
 & 1&2(2)$^*$     & $P$   \\
$\Phi_4$  & 1&23(2)             & $C$   &  $\Phi_9$
 & 1&6(1)         & $C$    \\
$\Phi_5$  & 1&28(2)             & $R$   & & \multicolumn{2}{c}{\mbox{}} & \\
\hline
\hline
\multicolumn{8}{|c|}{$L^2\cdot T=24^3,\qquad \beta_H=0.3438$}  \\
\hline
No.  &  \multicolumn{2}{c}{$aM[2^{++}]$} & Ops.   &
No.  &  \multicolumn{2}{c}{$aM[2^{++}]$} & Ops.   \\
\hline
$\Phi_1$  & 0&570(5)\er{0}{4}   & $P$ &  $\Phi_6$ 
 & 1&29(2)(2)       & $T$     \\
$\Phi_2$  & 0&791(9)            & $L$ &  $\Phi_7$ 
 & 1&31(3)\er{3}{0} & $R,L,C$ \\
$\Phi_3$  & 1&01(2)             & $R$ &  $\Phi_8$ 
 & 1&47$^*$         & $C$     \\
$\Phi_4$  & 1&09(2)             & $P$ &  $\Phi_9$ 
 & 1&49(4)          & $R$     \\
$\Phi_5$  & 1&1(1)$^*$          & $C$ & & \multicolumn{2}{c}{\mbox{}} & \\
\hline
\hline
\multicolumn{8}{|c|}{$L^2\cdot T=16^2\cdot24,\qquad \beta_H=0.3438$}  \\
\hline
No.  &  \multicolumn{2}{c}{$aM[2^{++}]$} & Ops.   &
No.  &  \multicolumn{2}{c}{$aM[2^{++}]$} & Ops.   \\
\hline
$\Phi_1$  & 0&361(12)\er{6}{0}  & $P$ &  $\Phi_6 $
 & 1&5(1)$^*$     & $C$   \\
$\Phi_2$  & 0&843(27)           & $T$ &  $\Phi_7$ 
 & 1&4(1)$^*$     & $L,R$ \\
$\Phi_3$  & 0&86(2)             & $L$ &  $\Phi_8$ 
 & 1&55(4)        & $C,T$ \\
$\Phi_4$  & 0&98(4)\er{0}{1}    & $P$ &  $\Phi_9$ 
 & 1&6(2)$^*$     & $L$   \\
$\Phi_5$  & 1&18(5)\er{3}{0}    & $C$ & $\Phi_{10}$ 
 & 1&79(7)        & $C$   \\
\hline
\hline
\end{tabular}
\caption{ \label{m2_lsmall}
{\em Mass spectrum and dominant contributions from the
operator basis in the $2^{++}$ channel at small scalar self-coupling
in the symmetric phase.}}
\end{center}
\end{table}
\begin{table}
\begin{center}
\begin{tabular}{|c|r@{.}l|r@{.}l|r@{.}l|}
\hline
\hline
 \multicolumn{7}{|c|}{$\beta_H=0.3438$}  \\
\hline
     &  \multicolumn{2}{c|}{$L^2\cdot T=36^3$} 
     &  \multicolumn{2}{c|}{$L^2\cdot T=24^3$}  
     &  \multicolumn{2}{c|}{$L^2\cdot T=16^2\cdot24$}  \\
\hline
No.  &  \multicolumn{2}{c|}{$aM[1^{--}]$} 
     &  \multicolumn{2}{c|}{$aM[1^{--}]$} 
     &  \multicolumn{2}{c|}{$aM[1^{--}]$} \\
$\Phi_1$  & 0&595(5)\er{0}{6}   & 0&582(14)            & 0&585(15)\err{17}{0}\\
$\Phi_2$  & 0&857(8)\err{12}{0} & 0&863(9)\er{0}{7}    & 0&99(2)$^*$  \\
$\Phi_3$  & 0&98(8)$^*$         & 1&068(18)\err{0}{14} & 1&15(4)      \\
\hline
\hline
\end{tabular}
\caption{ \label{m1_lsmall}
{\em Mass spectrum in the $1^{--}$ channel at small scalar
self-coupling in the symmetric phase.}} 
\end{center}
\end{table}

The physical states consist of a dense spectrum of bound states. The
second excited state in the $0^{++}$-channel is a pure gauge
excitation which, in this phase, cannot be perturbatively related to a
scattering state of two spin-1 particles. Instead it was suggested in
\cite{us} to interpret it as a glueball in analogy to the glueballs of
pure gauge theory. This interpretation is based on the observation
that the state is composed predominantly from gauge degrees of freedom
and that its mass nearly equals that of the lightest scalar glueball
in the pure SU(2) gauge theory at this value of $\beta_G$. As
Table\,\ref{m2_lsmall} shows, there is a corresponding state in the
$2^{++}$ channel. We also find higher excitations of these lowest
glueball states for both quantum numbers.

In Table \ref{PLcomp} we compare some properties of the Polyakov loop
operator in the Higgs and confinement regions. As explained in
subsection\,\ref{SECop}, the Polyakov loop has a large expectation
value in the Higgs phase, indicating the absence of confinement.
Similar to the correlator of Wilson loops, the perturbative expansion
of the Polyakov loop correlator for weak coupling is dominated by a
two-$W$ exchange diagram, and indeed we find its exponential fall-off
to be compatible with twice the $W$-mass. These findings are supported
by the observation that the mass extracted from the correlation of
Polyakov loops is roughly independent of the spatial lattice
size\,$L$, in sharp contrast to situations where confinement is
observed.

In the symmetric phase, on the other hand, we find that the
expectation value of the Polyakov loop is consistent with zero,
indicating confinement. Here, ordinary weak coupling expansions fail,
and the exponential fall-off of the correlation of Polyakov loops is
incompatible with twice the mass of the spin-1 state, c.f.~Table
\ref{PLcomp}. Instead, one interprets it as the mass of a loop of
chromoelectric flux, $aM_P$, which can be related to the string
tension $\sigma_L$
\beq
\label{string}
  \sum_{\vec{x}}\,\left\langle P_j^{(L)}(x)P_j^{(L)\dag}(0)
                  \right\rangle \simeq e^{-aM_P(L)t},\quad
   aM_P(L)=a^2\sigma_L L.
\eeq
Here $L$ is the spatial length of the lattice and hence of the flux
loop. An estimate for the string tension in infinite volume is then
provided by the relation\,\cite{for85}
\beq
\label{stringinfty}
  a^2\sigma_\infty = a^2\sigma_L+\frac{\pi}{6L^2}.
\eeq
We add that our numerical values for the string tension
$a\sqrt{\sigma_\infty}$ shown in Table\,\ref{PLcomp} are only around
3\% smaller than in the pure gauge theory\,\cite{tep93}.

\begin{table}[tbp]
\begin{center}
\begin{tabular}{||l||r@{.}l|r@{.}l|r@{.}l||r@{.}l|r@{.}l||}
\hline
\hline
  & \multicolumn{6}{c||}{Confinement}
  & \multicolumn{4}{c||}{Higgs}  \\
  & \multicolumn{6}{c||}{$\be_H=0.3438$}
  & \multicolumn{4}{c||}{$\be_H=0.3450$}  \\
\hline
$L^2\cdot T$ & \multicolumn{2}{c|}{$36^3$}
             & \multicolumn{2}{c|}{$24^3$}
             & \multicolumn{2}{c||}{$16^2\cdot24$} 
             & \multicolumn{2}{c|}{$16^2\cdot24$}
             & \multicolumn{2}{c||}{$24^3$} \\
\hline
\hline
$\langle P_1^{(L)}\rangle$
             & 0&001(10)  & 0&001(10) & 0&14(2)
             & 6&535(6)   & 6&724(5)  \\
$aM_{P}$     & 0&879(6)\er{0}{2} & 0&570(4)  & 0&366(3)
             & 1&67(4)    & 1&54(4)${}^*$ \\
$aM[1^{--}]$ & 0&595(5)\er{0}{6} & 0&582(14) & 0&585(15)\err{17}{0}
             & 0&839(4)   & 0&837(3)\er{0}{4} \\
$a\sqrt{\sigma_\infty}$
             & 0&1575(5)\er{0}{2} & 0&1570(6)\er{0}{2} & 0&1579(5)
             & \multicolumn{2}{c|}{\mbox{--}}
             & \multicolumn{2}{c||}{\mbox{--}} \\
\hline
\hline
\end{tabular}
\caption{\label{PLcomp} {\em Properties of the Polyakov loop operator
in the confinement and Higgs region for $\be_G=9$. 
$\langle P_1^{(L)}\rangle$ here refers to the unsmeared operator.}}
\end{center}
\end{table}

\subsection{Finite-volume effects and toroidal operators}

In this subsection we consider the issue of finite-size effects in
more detail. In particular, we want to clarify the r\^ole of toroidal
(or winding) operators, i.e. operators constructed out of Polyakov
loops, for finite-volume studies in QCD-like situations. This is of
particular relevance in the symmetric phase of our model, where we
have reported strong finite-size effects in the ground state of the
$0^{++}$ channel\,\cite{us}. Although we include this discussion in
the section on small scalar coupling, our findings are relevant at
large coupling as well.

Toroidal operators, such as the Polyakov loop, project mainly onto
states encircling the periodic boundary conditions, sometimes called
``torelons''\,\cite{cm87a} (see also eq.\,(\ref{string})). These torelon
states can interfere with the mass spectrum of the theory in infinite
volume if
\beq
\label{EQfinvol}
a^2\sigma_L L \leqsim aM,
\eeq
where $M$ is the mass of an eigenstate of the Hamiltonian. It should
be stressed at this point that torelon states exist in the continuum
limit for finite box size and periodic boundary conditions. Thus they
are not lattice artefacts, i.e. artefacts arising from the
discretisation of the theory.

As explained in subsection\,\ref{SECop}, we use blocked operators in
the $0^{++}$ and $2^{++}$ channels, which, due to the iterative nature
of the blocking procedure, may contain contributions which wind around
the lattice and can thus have a sizeable projection onto torelon
states. If relation\,(\ref{EQfinvol}) applies this can then lead to a
possible misidentification of the ground state and/or higher
excitations of the Hamiltonian. This scenario does not apply in the
$1^{--}$ channel, where it is impossible to construct a torelon
operator with the same quantum numbers. Furthermore, the overlap
between single torelons and glueball operators will be completely
suppressed in the pure gauge theory where we have confinement and the
associated $Z_2$ symmetry, under which torelon and glueball operators
transform differently. However, torelon-antitorelon pairs can have a
big influence on the spectrum of glueballs and are therefore included
in our study.

\begin{figure}[th]
\begin{center}
\leavevmode
\epsfysize=250pt
\epsfbox[20 30 620 730]{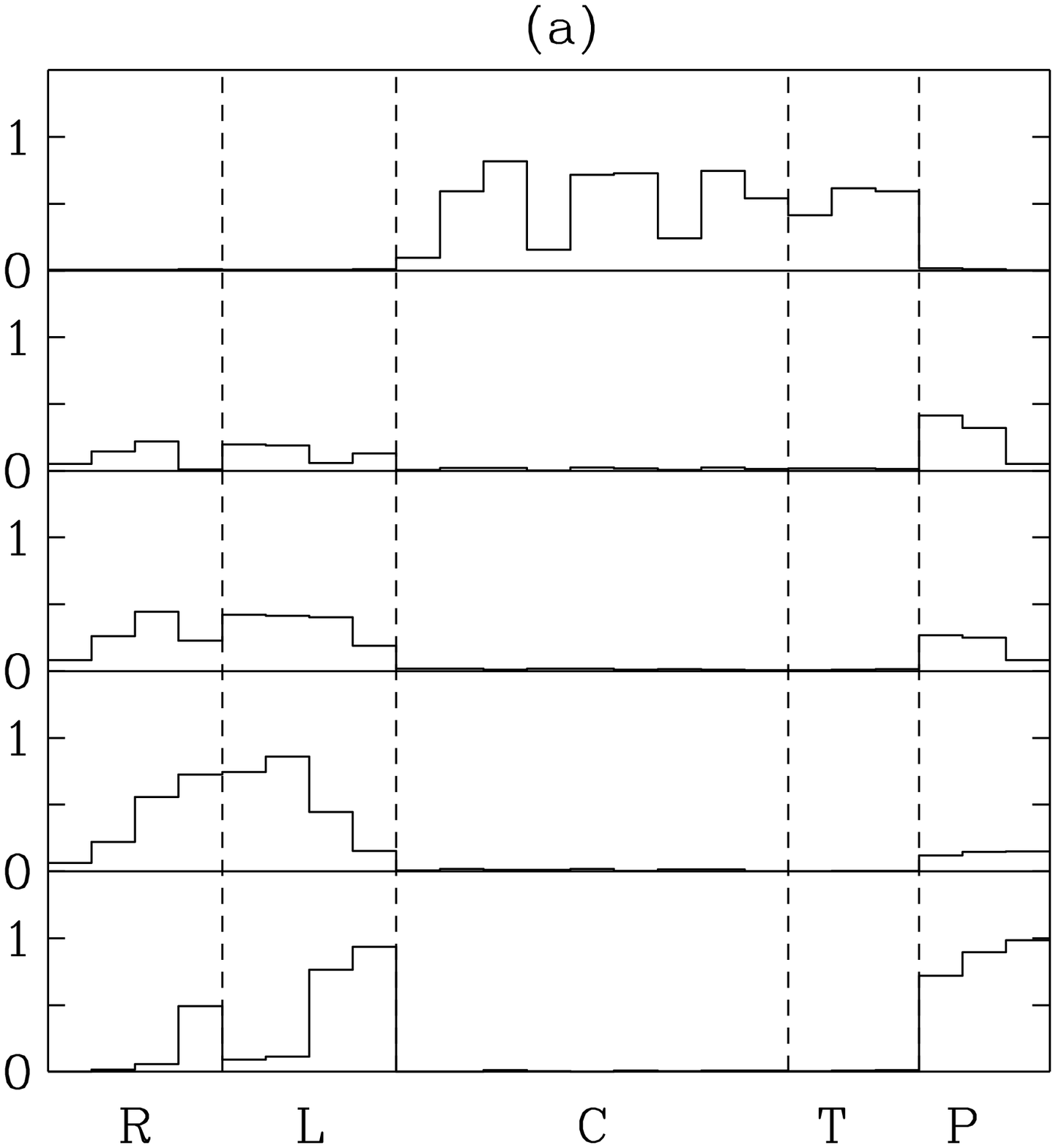}
%
\leavevmode
\epsfysize=250pt
\epsfbox[20 30 620 730]{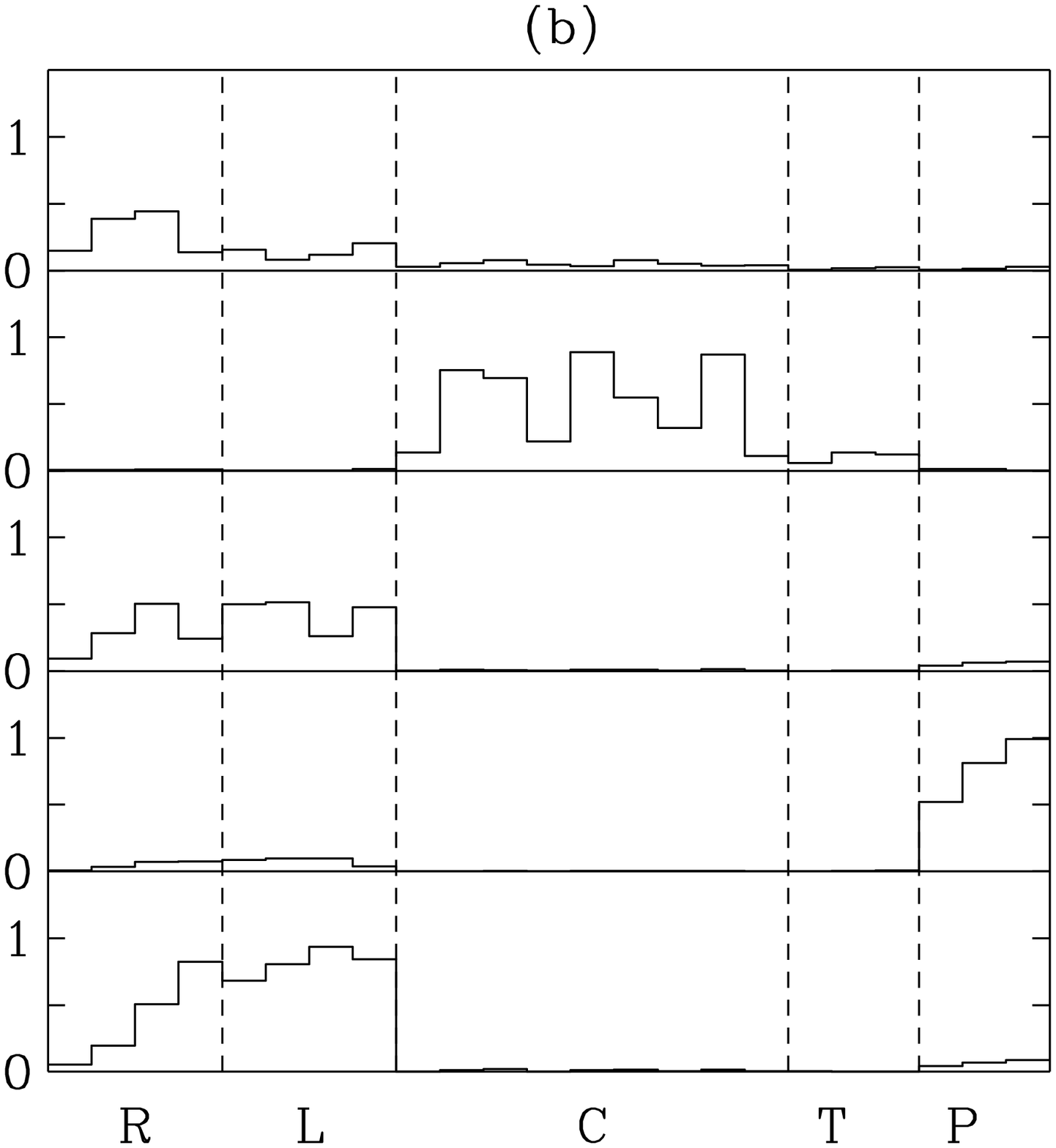}
\end{center}
\vspace{-1.6cm}
\caption[]{\it \label{op2}
The coefficients $a_{ik}$ of the 23 operators used in the simulation
    for the five lowest $2^{++}$ eigenstates, (a) $L=24$,
(b) $L=36$.}
\end{figure}

As an illustration of the issue, consider Fig.~\ref{spec_lsmall} (b)
and Table\,\ref{m0_lsmall}. The analysis of the operator content shows
$\Phi_4$ to have almost exclusively $P$-content. The mass associated
with it is a flux loop mass appropriate to extract the string tension,
but it is not part of the spectrum of the Hamiltonian in infinite
volume and has to be removed. An example how these toroidal states mix
with the physical ones and have to be carefully disentangled is given
for the case of the $2^{++}$ spectrum in Fig.~\ref{op2} and
Table\,\ref{m2_lsmall}.
On the smaller lattice, Fig.~\ref{op2}\,(a), the lowest state is a
mixture of $R/L$ and $P$-type operators, with a slight dominance by
$P$. Quoting the corresponding mass as a physical state close to its
infinite-volume limit would be erroneous, as is elucidated by
considering the same operators on a larger volume. The masses arising
from winding operators have become heavier (see
eq.\,(\ref{EQfinvol})), while the non-winding operators produce mass
values which hardly change under variation of the volume, and are thus
already close to infinite-volume physics. Combining the information
about the overlaps of individual operators with their behaviour as the
spatial lattice size is increased makes the identification of the
masses which are part of the spectrum in the infinite-volume limit
unambiguous.

As a result, some of the conclusions concerning the finite-volume
behaviour in ref.\,\cite{us} have to be revised. First, the large
finite-volume effects in the $0^{++}$ channel clearly arise from
toroidal operators, as the relation\,(\ref{EQfinvol}) certainly holds
in the $0^{++}$ channel for the smallest lattices considered in our
previous work. Second, our new findings remove the anomalous behaviour
of the first excited state. This can be understood since the ordering
of the excitation spectrum changes with increasing lattice size, as
toroidal operators will produce more and more massive states.

\subsection{The physical spectrum}

After removing finite size effects introduced by considering toroidal
operators, we are ready to present our final mass estimates for the
spectrum in Table \ref{mfin_lsmall}. In addition to the mass spectrum
extracted from our simulations of the SU(2) Higgs model, the glueball
spectrum of the pure SU(2) gauge theory is also
given\,\cite{tepunp,tep93}. 

For all cases considered we find that the glueball masses deviate from
those in the pure gauge theory at the percent level at most. Thus, the
glueball spectrum of the pure gauge theory is almost identically
repeated in the scalar-gauge theory and appears to be entirely
insensitive to the presence of the scalar fields. In addition, and
apparently disjoint from this part of the spectrum, there are bound
states of scalars which also have some gauge content. Combining this
observation with the practically identical value of the string
tension, we confirm our earlier conclusion\,\cite{us} that the pure
gauge sector of the theory decouples almost completely from the scalar
sector at this point in the symmetric phase.
\begin{table}
\begin{center}
\begin{tabular}{||r@{.}lr@{.}l|r@{.}l||r@{.}lr@{.}l|r@{.}l||}
\hline
\hline
\multicolumn{6}{||c||}{$0^{++}$ channel} & 
\multicolumn{6}{c||}{$2^{++}$ channel} \\
\hline
\multicolumn{4}{||c|}{Gauge-Higgs} & \multicolumn{2}{c||}{Pure Gauge} &
\multicolumn{4}{c|}{Gauge-Higgs} & \multicolumn{2}{c||}{Pure Gauge} 
\\
\hline
\multicolumn{2}{||c}{Scalar} & \multicolumn{2}{c|}{glueball} & 
\multicolumn{2}{c||}{glueball} &
\multicolumn{2}{|c}{Scalar} & \multicolumn{2}{c|}{glueball} & 
\multicolumn{2}{c||}{glueball} \\
\hline
0&379(5)\er{0}{7} &  \mcemptyr   &  \mcemptyrr  &
0&764(7)          &  \mcemptyr   &  \mcemptyrr  \\
0&706(5)\er{0}{5} &  \mcemptyr   &  \mcemptyrr  &
0&99(3)           &  \mcemptyr   &  \mcemptyrr  \\
\mcemptyll        & 0&74(3)$^*$  &  0&767(6)    &
\mcemptyl         & 1&23(2)      &  1&26(2)     \\
0&95(3)           &  \mcemptyr   &  \mcemptyrr  &
1&28(2)           &  \mcemptyr   &  \mcemptyrr  \\
0&92(3)\er{2}{0}  &  \mcemptyr   &  \mcemptyrr  &
1&25(15)$^*$      &  \mcemptyr   &  \mcemptyrr  \\
\mcemptyll        & 1&08(4)\er{2}{0} &  1&08(2) &
\mcemptyl         & 1&46(2)          &  1&50(5) \\
\mcemptyll        & 1&35(2)          &  1&27(2) &
\mcemptyl         & 1&64(4)$^*$      &  1&77(6) \\
\mcemptyll        & \mcemptyr        &  \mcemptyrr &
1&67(6)           & \mcemptyr        &  \mcemptyrr  \\
\hline
\hline
\end{tabular}
\caption{ \label{mfin_lsmall}
{\em Final mass estimates using the
data obtained on $36^3$ at $\beta_G=9$, $\beta_H=0.3438$ in the
symmetric phase at small scalar self-coupling. Our data for the
glueball are compared to results obtained in the pure gauge theory.}}
\end{center}
\end{table}

\section{Results at large scalar coupling}
\label{SEClarge}
It is now interesting to study whether a significant increase of the
scalar self coupling destroys the decoupling of the pure gauge sector
observed for small coupling. For this purpose we fix the ratio of
scalar and gauge couplings to be about ten times larger than
previously, $\lambda_3/g_3^2=0.2743$. At this parameter value the
first order phase transition between the Higgs and the confinement
regions upon variation of $\be_H$ has disappeared and turned into a
smooth crossover \cite{bp94,kaj96}. One would expect that the absence of a
phase transition weakens the specific Higgs and confinement properties
on the respective ``sides" of the crossover. For the simulations the
points $\be_H=0.35262$ and $\be_H=0.35542$ are chosen to represent the
confinement and Higgs sides of the phase diagram, respectively.

\begin{figure}[tbp]
\begin{center}
\leavevmode
\epsfysize=250pt
\epsfbox[20 30 620 730]{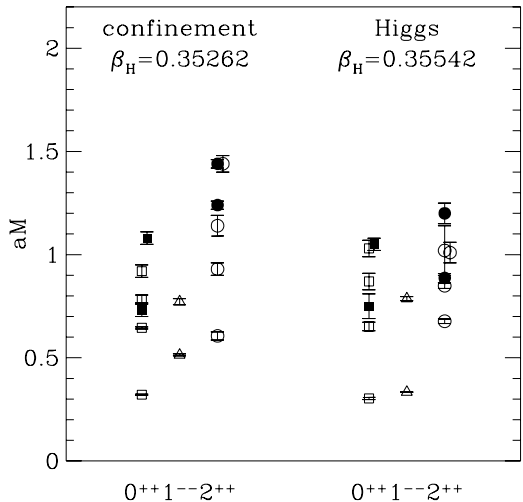}
\end{center}
\vspace{-1.6cm}
\caption[]{\it \label{spec_llarge}
 The lowest states of the spectrum in the confinement (left) and Higgs
 (right) region. Full symbols indicate states that receive
 predominantly contributions from Wilson loops.}
\end{figure}

The lowest states of the mass spectrum are shown in
Fig.~\ref{spec_llarge}. Higher excitations as well as the composition
of all states can be found in Tables \ref{m0s_llarge}-\ref{m1_llarge}.
The spectrum in the confinement region looks qualitatively the same as
in the small scalar coupling case. In particular, we observe the same
occurrence of glueball states in the $0^{++}$ and $2^{++}$ channels as
well as their non-mixing with the bound states of scalars as for small
scalar coupling, c.f.~Tables \ref{m0s_llarge}, \ref{m2s_llarge}. By
comparing the numerical values of the glueball masses at small and
large scalar coupling, one finds that all glueball masses agree
quantitatively, i.e. are statistically compatible, with those at small
scalar coupling. Thus the previously stated decoupling of the pure
gauge sector from the Higgs part of the theory extends over an order
of magnitude increase in the ratio of scalar coupling to gauge
coupling!

\begin{table}[tbp]
\begin{center}
\begin{tabular}{|c|r@{.}lc|c|r@{.}lc|}
\hline
\hline
\multicolumn{8}{|c|}{$L^2\cdot T=36^3,\qquad \beta_H=0.3526$}  \\
\hline
No.  &  \multicolumn{2}{c}{$aM[0^{++}]$} & Ops.   & 
No.  &  \multicolumn{2}{c}{$aM[0^{++}]$} & Ops.   \\
\hline
$\Phi_1$  & 0&321(3)          & $R,L$ &  $\Phi_8$ 
 & 1&06(6)$^*$    & $S$   \\
$\Phi_2$  & 0&645(5)\er{0}{3} & $R,L$ &  $\Phi_9$ 
 & 1&33(3)(1)     & $C$   \\
$\Phi_3$  & 0&73(3)           & $C$   & $\Phi_{10}$ 
 & 1&21(9)$^*$    & $L$   \\
$\Phi_4$  & 0&850(7)\er{3}{0} & $P$   & $\Phi_{11}$ 
 & 1&40(3)$^*$    & $P$  \\
$\Phi_5$  & 0&785(19)         & $L$   & $\Phi_{12}$ 
 & 1&56(4)$^*$    & $C$   \\
$\Phi_6$  & 0&92(3)\er{0}{2}  & $R,L$ & $\Phi_{13}$ 
 & 1&60(4)        & $C$   \\
$\Phi_7$  & 1&08(3)\er{0}{2}  & $C$   & $\Phi_{14}$ 
 & 1&74(5)        & $T$   \\
\hline
\hline
\multicolumn{8}{|c|}{$L^2\cdot T=24^3,\qquad \beta_H=0.3526$}  \\
\hline
No.  &  \multicolumn{2}{c}{$aM[0^{++}]$} & Ops.   & 
No.  &  \multicolumn{2}{c}{$aM[0^{++}]$} & Ops.   \\
\hline
$\Phi_1$  & 0&324(4)          & $L$      &  $\Phi_9$ 
 & 1&10(3)     &       \\
$\Phi_2$  & 0&543(9)          & $P$      & $\Phi_{10}$ 
 & 1&17(3)     &  $C,T$  \\
$\Phi_3$  & 0&661(7)          & $R,L,P$  & $\Phi_{11}$ 
 & 1&18(3)     &  $C,T$  \\
$\Phi_4$  & 0&737(12)         & $C$      & $\Phi_{12}$ 
 & 1&42(3)$^*$ &  $T$    \\
$\Phi_5$  & 0&752(23)\err{0}{26}& $L$    & $\Phi_{13}$ 
 & 1&3(2)$^*$  &  $L$    \\
$\Phi_6$  & 0&968(12)         & $R,L$    & $\Phi_{14}$ 
 & 1&53(3)     &  $C,T$  \\
$\Phi_7$  & 0&98(3)           & $C,P_d$  & $\Phi_{15}$ 
 & 1&53(4)     &  $C,T$  \\
$\Phi_8$  & 1&03(2)           & $C,P_d$  & &  \multicolumn{2}{c}{\mbox{}} & \\
\hline
\hline
\end{tabular}
\caption{ \label{m0s_llarge}
{\em Mass spectrum and dominant contributions from the
operator basis in the $0^{++}$ channel at large scalar self-coupling
in the symmetric phase.}} 
\end{center}
\end{table}
\begin{table}
\begin{center}
\begin{tabular}{|c|r@{.}lc|c|r@{.}lc|}
\hline
\hline
\multicolumn{8}{|c|}{$L^2\cdot T=36^3,\qquad \beta_H=0.3526$}  \\
\hline
No.  &  \multicolumn{2}{c}{$aM[2^{++}]$} & Ops.   & 
No.  &  \multicolumn{2}{c}{$aM[2^{++}]$} & Ops.   \\
\hline
$\Phi_1$  & 0&606(20)\err{11}{0} & $R,L$ &  $\Phi_7$
 & 1&45(3)     & $P$    \\
$\Phi_2$  & 0&859(17)\err{0}{15} & $P$   &  $\Phi_8$
 & 1&44(4)$^*$ & $R$     \\
$\Phi_3$  & 0&93(3)$^*$          & $R,L$ &  $\Phi_9$
 & 1&62(4)     & $C$     \\
$\Phi_4$  & 1&24(2)$^*$          & $C$   & $\Phi_{10}$
 & 1&2(2)$^*$  & $L$     \\
$\Phi_5$  & 1&14(5)$^*$          & $L$   & $\Phi_{11}$
 & 1&6(1)$^*$  & $C,T$   \\
$\Phi_6$  & 1&44(2)              & $C$   & $\Phi_{12}$
 & 1&81(4)     & $C,T$   \\
\hline
\hline
\multicolumn{8}{|c|}{$L^2\cdot T=24^3,\qquad \beta_H=0.3526$}  \\
\hline
No.  &  \multicolumn{2}{c}{$aM[2^{++}]$} & Ops.   & 
No.  &  \multicolumn{2}{c}{$aM[2^{++}]$} & Ops.   \\
\hline
$\Phi_1$  & 0&568(5)\er{3}{0}    & $P$   &  $\Phi_6$
 & 1&29(2)\er{0}{1} & $C,T$  \\
$\Phi_2$  & 0&734(23)            & $L$   &  $\Phi_7$
 & 1&22(2)\er{0}{2} & $L$    \\
$\Phi_3$  & 1&02(2)\er{0}{1}     & $R$   &  $\Phi_8$
 & 1&48(4)\er{0}{2} & $C$    \\
$\Phi_4$  & 1&12(2)              & $P$   &  $\Phi_9$
 & 1&40(15)$^*$     & $R$    \\
$\Phi_5$  & 1&22(3)              & $C,T$ & $\Phi_{10}$
 & 1&65(5)          & $R$    \\
\hline
\hline
\end{tabular}
\caption{\label{m2s_llarge}
{\em Mass spectrum and dominant contributions from the
operator basis in the $2^{++}$ channel at large scalar self-coupling
in the symmetric phase.}} 
\end{center}
\end{table}

On the Higgs side of the crossover, a dramatic change in the nature of
the spectrum has taken place. The specific ``Higgs-like'' feature of
having only the Higgs and $W$-particle as low lying states with a
large gap to higher excitations has entirely disappeared. Instead, we
now have a dense spectrum of states on the Higgs side as well, which
differs only quantitatively from the one on the confinement side. Note
that the value for the scalar coupling considered here is so large
that perturbation theory is not reliable anymore \cite{bfh}.

\begin{table}
\begin{center}
\begin{tabular}{|c|r@{.}lc|c|r@{.}lc|}
\hline
\hline
\multicolumn{8}{|c|}{$L^2\cdot T=36^3,\qquad \beta_H=0.3554$}  \\
\hline
No.  &  \multicolumn{2}{c}{$aM[0^{++}]$} & Ops.   & 
No.  &  \multicolumn{2}{c}{$aM[0^{++}]$} & Ops.   \\
\hline
$\Phi_1$  & 0&303(5)\er{4}{1}    & $R,L$         &  $\Phi_9$
 & 1&02(4)   & $(P,T,P_d)$ \\
$\Phi_2$  & 0&652(21)\err{0}{15} & $L,C,P,T,P_d$ & $\Phi_{10}$
 & 1&33(2)   & $C$    \\
$\Phi_3$  & 0&80(4)              & $P,T,P_d$     & $\Phi_{11}$
 & 1&32(2)   & $C$    \\
$\Phi_4$  & 0&75(6)$^*$          & $C$           & $\Phi_{12}$
 & 1&39(4)\er{0}{2} & $P_d$ \\
$\Phi_5$  & 0&75(5)$^*$          & $C,(P)$       & $\Phi_{13}$
 & 1&41(3)   & $C,T$  \\
$\Phi_6$  & 1&05(3)\er{0}{1}     & $C$           & $\Phi_{14}$
 & 1&51(3)\er{0}{1} & $C,T$ \\
$\Phi_7$  & 0&87(4)\er{0}{1}$^*$ &               & $\Phi_{15}$
 & 1&53(4)   & $C,T$  \\
$\Phi_8$  & 1&03(4)$^*$          &             & & \multicolumn{2}{c}{\mbox{}} & \\
\hline
\hline
\multicolumn{8}{|c|}{$L^2\cdot T=24^3,\qquad \beta_H=0.3554$}  \\
\hline
No.  &  \multicolumn{2}{c}{$aM[0^{++}]$} & Ops.   & 
No.  &  \multicolumn{2}{c}{$aM[0^{++}]$} & Ops.   \\
\hline
$\Phi_1$  & 0&317(3)\er{0}{6}    & $L$    &  $\Phi_7$
 & 1&05(5)$^*$    & $C$  \\
$\Phi_2$  & 0&642(14)\err{0}{16} & $L,P$  &  $\Phi_8$
 & 1&20(2)        & $C$  \\
$\Phi_3$  & 0&704(24)            & $P,T$  &  $\Phi_9$
 & 1&21(3)        & $C,P,T$ \\
$\Phi_4$  & 0&82(5)              & $L$    & $\Phi_{10}$
 & 1&30(3)        & $T,(C)$  \\
$\Phi_5$  & 0&87(2)$^*$          & $C$    & $\Phi_{11}$
 & 1&37(3)        & $P,T$   \\
$\Phi_6$  & 0&9(1)$^*$           &        & $\Phi_{12}$
 & 1&37(4)        & $C,T$    \\
\hline
\hline
\end{tabular}
\caption{ \label{m0h_llarge}
{\em Mass spectrum and dominant contributions from the
operator basis in the $0^{++}$ channel at large scalar self-coupling
in the broken phase. Labels in brackets indicate that the contribution
from the corresponding operator are small but significant.}} 
\end{center}
\end{table}

\begin{table}
\begin{center}
\begin{tabular}{|c|r@{.}lc|c|r@{.}lc|}
\hline
\hline
\multicolumn{8}{|c|}{$L^2\cdot T=36^3,\qquad \beta_H=0.3554$}  \\
\hline
No.  &  \multicolumn{2}{c}{$aM[2^{++}]$} & Ops.   & 
No.  &  \multicolumn{2}{c}{$aM[2^{++}]$} & Ops.   \\
\hline
$\Phi_1$  & 0&677(11)      & $L$       &  $\Phi_7$
 & 1&2(1)$^*$  & $L$   \\
$\Phi_2$  & 0&849(13)      & $R,L,P,T$ &  $\Phi_8$
 & 1&43(4)     & $C$   \\
$\Phi_3$  & 0&887(22)      & $(C)$     &  $\Phi_9$
 & 1&44(2)     & $C$   \\
$\Phi_4$  & 1&02(12)       & $(R,L)$   & $\Phi_{10}$
 & 1&57(4)     & $L$   \\
$\Phi_5$  & 1&01(5)\er{0}{3}  & $(R)$  & $\Phi_{11}$
 & 1&51(3)\er{1}{0} & $T$ \\
$\Phi_6$  & 1&20(5)\er{0}{2}  & $C$    & & \multicolumn{2}{c}{\mbox{}} & \\
\hline
\hline
\multicolumn{8}{|c|}{$L^2\cdot T=24^3,\qquad \beta_H=0.3554$}  \\
\hline
No.  &  \multicolumn{2}{c}{$aM[0^{++}]$} & Ops.   & 
No.  &  \multicolumn{2}{c}{$aM[0^{++}]$} & Ops.   \\
\hline
$\Phi_1$  & 0&639(8)\err{0}{22} & $L$   &  $\Phi_6$
 & 1&10(7)    & $C$   \\
$\Phi_2$  & 0&84(3)$^*$         & $R,L$ &  $\Phi_7$
 & 1&11(8)    & $T$   \\
$\Phi_3$  & 0&87(4)             & $C$   &  $\Phi_8$
 & 1&34(3)    & $C$   \\
$\Phi_4$  & 0&95(10)$^*$        & $R$   &  $\Phi_9$
 & 1&34(2)\er{0}{2} & $C$ \\
$\Phi_5$  & 1&21(3)             & $L$   & $\Phi_{10}$
 & 1&46(6)    &     \\
\hline
\hline
\end{tabular}
\caption{ \label{m2h_llarge}
{\em Mass spectrum and dominant contributions from the
operator basis in the $2^{++}$ channel at large scalar self-coupling
in the broken phase. Labels in brackets indicate that the contribution
from the corresponding operator are small but significant.}} 
\end{center}
\end{table}
\begin{table}
\begin{center}
\begin{tabular}{|c|r@{.}l|r@{.}l||r@{.}l|r@{.}l|}
\hline
\hline
 & \multicolumn{4}{c||}{$\beta_H=0.3526$}  
 & \multicolumn{4}{c|}{$\beta_H=0.3554$}  \\
\cline{2-9}
     &  \multicolumn{2}{c|}{$L^2\cdot T=36^3$} 
     &  \multicolumn{2}{c||}{$L^2\cdot T=24^3$}  
     &  \multicolumn{2}{c|}{$L^2\cdot T=36^3$} 
     &  \multicolumn{2}{c|}{$L^2\cdot T=24^3$}  \\
\hline
No.  &  \multicolumn{2}{c|}{$aM[1^{--}]$} 
     &  \multicolumn{2}{c||}{$aM[1^{--}]$} 
     &  \multicolumn{2}{c|}{$aM[1^{--}]$} 
     &  \multicolumn{2}{c|}{$aM[1^{--}]$} \\
\hline
$\Phi_1$  & 0&514(5)           & 0&482(15)\err{41}{10}
   & 0&3341(12)\err{0}{14} & 0&3313(27)  \\
$\Phi_2$  & 0&772(15)\er{6}{0} & 0&80(2)              
   & 0&787(10)             & 0&822(17)   \\
$\Phi_3$  & 0&94(4)\er{1}{0}   & 0&94(5)            
   & 1&040(16)             & 0&99(8)$^*$ \\
\hline
\hline
\end{tabular}
\caption{ \label{m1_llarge}
{\em Mass spectrum in the $1^{--}$ channel at large scalar
self-coupling in the symmetric and broken phases.}}
\end{center}
\end{table}

Let us now look at the expectation value of the Polyakov loop. The
points where we analysed the mass spectrum correspond to the endpoints
in Fig.~\ref{vev}, where the expectation value of the Polyakov loop is
shown. Note that now there is a small but non-zero, measurable value
for the expectation value of the Polyakov loop on the confinement
side. This is a first indication that colour charges are screened at
large distances. On the other hand, at the Higgs end we measure a
large expectation value, just as in the case of small scalar
coupling. Comparing the two sides in Fig.~\ref{spec_llarge} one might
be inclined to conclude that the similarity of the spectra indicates
an extension of the confining physics over to the Higgs side of the
crossover, which seems not unreasonable given the absence of a real
transition. However, from the large difference in the expectation
value for the Polyakov loop we are led to conclude that the dynamics
responsible for the dense spectrum on the Higgs side must be very
different from that on the confinement side. In particular, no flux
loops exist on this side of the crossover. This may be an indication
that what we see in this region of the phase diagram is a strong
scalar coupling Higgs regime. More numerical and analytic evidence is
required, though, to verify this interpretation.

Our final mass estimates for the physical states in the confinement
phase, with the toroidal operators removed as in the previous section,
are given in Table \ref{mfin_llarge}.
\begin{table}
\begin{center}
\begin{tabular}{||r@{.}lr@{.}l|r@{.}l||r@{.}lr@{.}l|r@{.}l||}
\hline
\hline
\multicolumn{6}{||c||}{$0^{++}$ channel} & 
\multicolumn{6}{c||}{$2^{++}$ channel} \\
\hline
\multicolumn{4}{||c|}{Gauge-Higgs} & \multicolumn{2}{c||}{Pure Gauge} &
\multicolumn{4}{c|}{Gauge-Higgs} & \multicolumn{2}{c||}{Pure Gauge} 
\\
\hline
\multicolumn{2}{||c}{Scalar} & \multicolumn{2}{c|}{glueball} & 
\multicolumn{2}{c||}{glueball} &
\multicolumn{2}{|c}{Scalar} & \multicolumn{2}{c|}{glueball} & 
\multicolumn{2}{c||}{glueball} \\
\hline
0&321(3)          &  \mcemptyr   &  \mcemptyrr  &
0&61(2)\er{1}{0}  &  \mcemptyr   &  \mcemptyrr  \\
0&645(5)\er{0}{3} &  \mcemptyr   &  \mcemptyrr  &
0&93(3)$^*$       &  \mcemptyr   &  \mcemptyrr  \\
\mcemptyll        & 0&73(3)      &  0&767(6)    &
\mcemptyl         & 1&24(2)$^*$  &  1&26(2)     \\
0&79(2)           &  \mcemptyr   &  \mcemptyrr  &
1&14(5)$^*$       &  \mcemptyr   &  \mcemptyrr  \\
0&92(3)\er{0}{2}  &  \mcemptyr   &  \mcemptyrr  &
1&44(4)$^*$       &  \mcemptyr   &  \mcemptyrr  \\
\mcemptyll        & 1&08(3)\er{0}{2} &  1&08(2) &
\mcemptyl         & 1&44(2)          &  1&50(5) \\
\mcemptyll        & 1&33(3)          &  1&27(2) &
\mcemptyl         & 1&62(4)          &  1&77(6) \\
\hline
\hline
\end{tabular}
\caption{ \label{mfin_llarge}
{\em Final mass estimates using the
data obtained on $36^3$ at $\beta_G=9$, $\beta_H=0.3526$ in the
symmetric phase at large scalar self-coupling. Our data for the
glueball are compared to results obtained in the pure gauge theory.}}
\end{center}
\end{table}

\section{The crossover region}

As we have seen in the last section, the mass spectra of the Higgs and
confinement regions are rather similar at large scalar coupling,
although the underlying dynamics must be quite different. Since there
is no phase transition but only a smooth crossover ``separating" the
respective regions at large scalar coupling one can study how the
spectra are continuously connected. Apart from mapping the spectra in
the Higgs and confinement regions onto each other, this also offers
the opportunity to follow the change in the dynamics as one moves from
one regime into the other.

\begin{figure}
\begin{center}
\leavevmode
\epsfysize=250pt
\epsfbox[20 30 620 730]{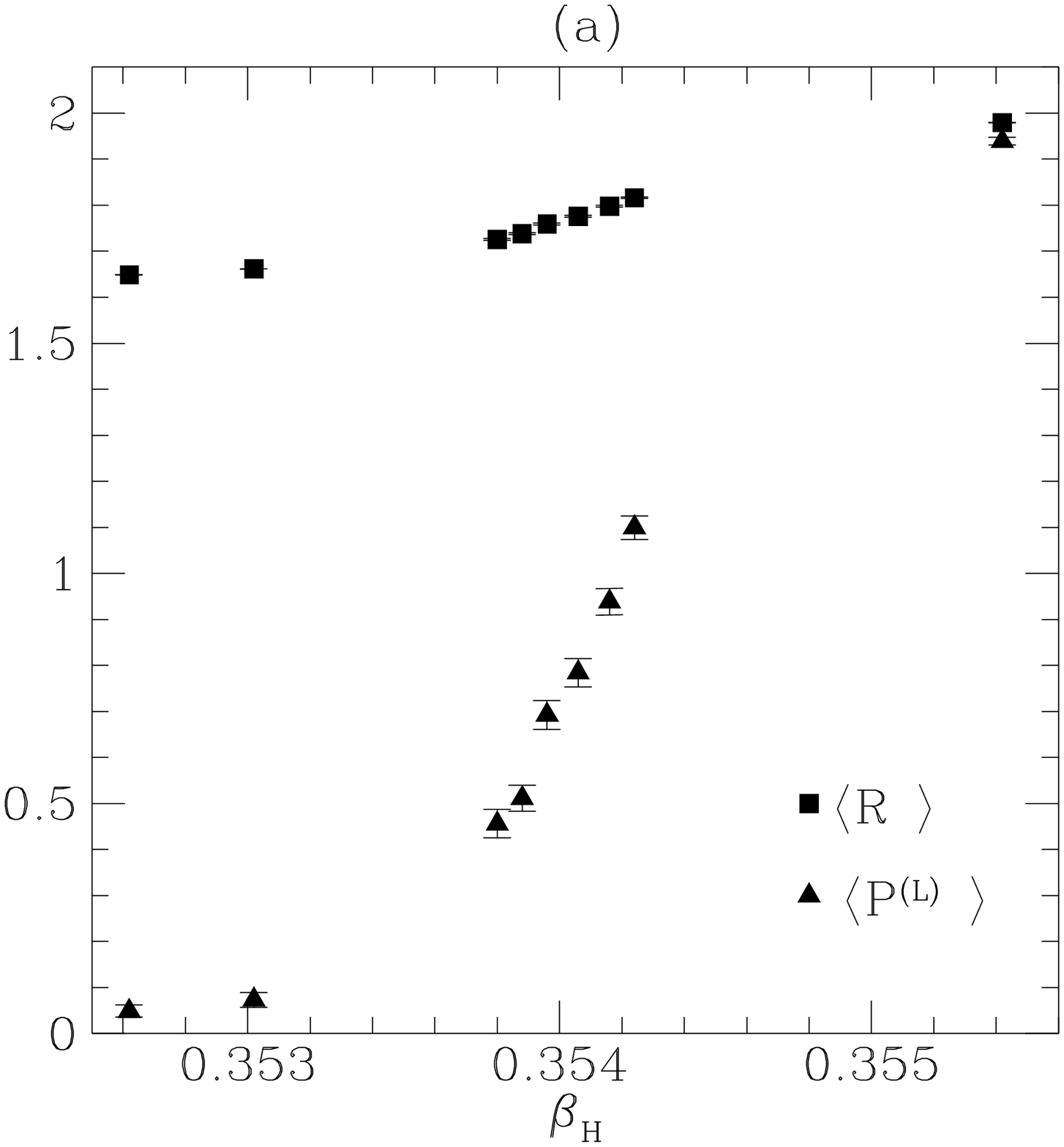}
\leavevmode
\epsfysize=250pt
\epsfbox[20 30 620 730]{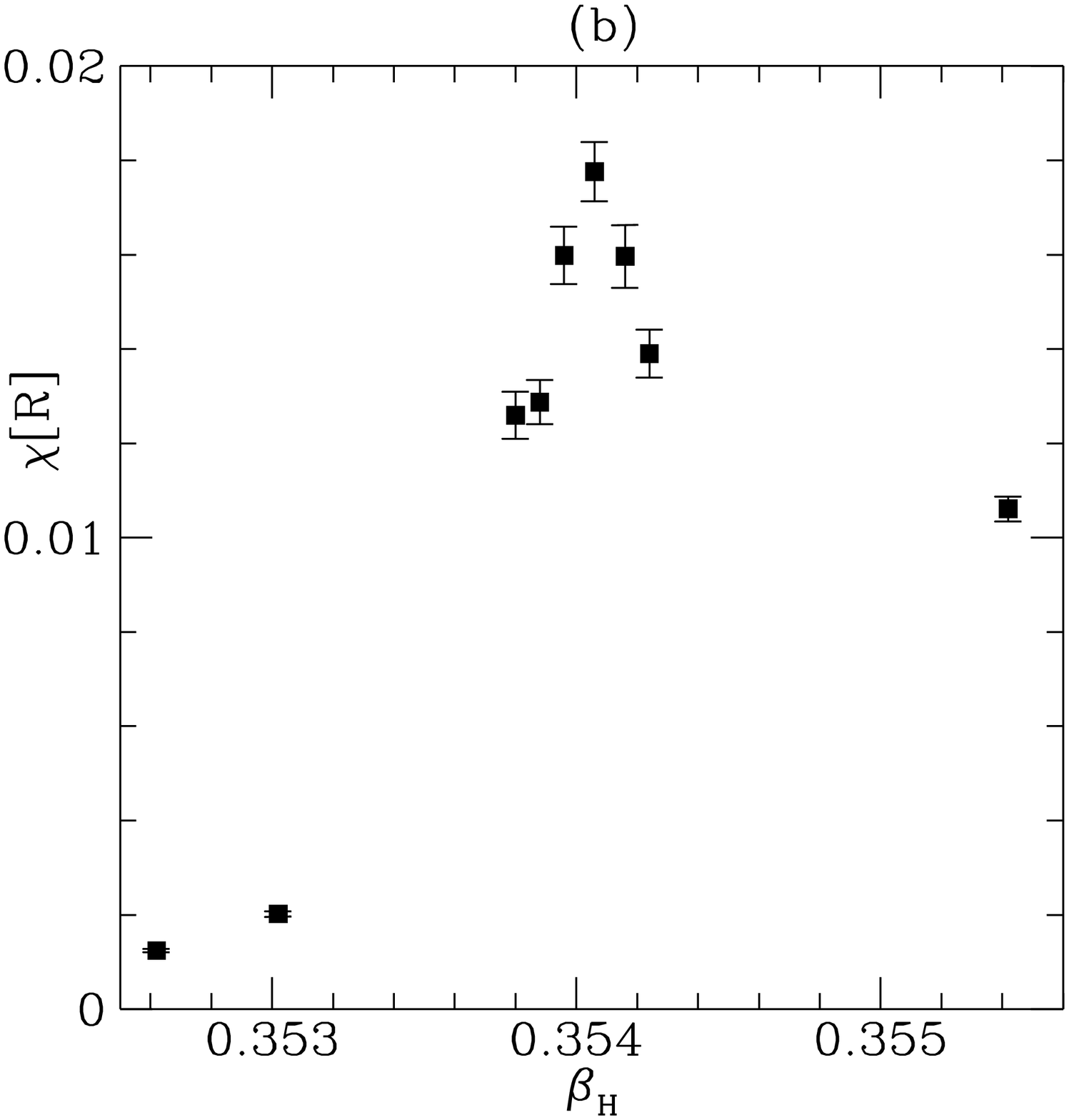}
\vspace{-1.6cm}
\end{center}
\caption[]{\it \label{vev}
 (a) Vacuum expectation values for $R$ and $P$.
 (b) Susceptibility of a smeared $R$-operator.}
\end{figure}
We start by considering the expectation values of the Polyakov loop
and the length of the Higgs field, displayed in Fig.~\ref{vev}.  In
most simulations dealing with the phase transition $R$ has been used
as an ``order parameter", as it jumps across a first order phase
transition from small values in the confinement phase to large values
in the Higgs phase. As Fig.~\ref{vev} illustrates, this jump has
entirely disappeared, the change in $\langle R\rangle$ is smooth and
very slow, with its values differing by only $\sim 30\%$ at the end
points studied in the last section. On the other hand, as mentioned
in the previous section, the expectation value of the Polyakov line is
quite different at both ends of the curve. Because it probes confining
behaviour and the change in dynamics in this region of parameter
space, the expectation value of the Polyakov loop is more sensitive to
the changes between the regions and hence makes a better ``order
parameter". However, strictly speaking it has to remain non-zero on
the confinement side as well, so the change is smooth and quantitative
only, as expected for a crossover.

Despite the smoothness of the crossover, it is possible to define a
critical hopping parameter through the peak of the susceptibility of
various quantities. We obtained the best signal for the susceptibility
of a smeared version of our $R$-operator,
\beq
 \chi= \langle R^2
\rangle - \langle R \rangle^2 \;.  
\eeq
In Fig.~\ref{vev}\,(b) one easily identifies a pronounced peak in the
susceptibility indicating large fluctuations in the operator under
consideration. As in ref.\, \cite{kaj96} this peak serves to locate
the critical value of the hopping parameter $\be^c_H$ at which the
crossover takes place, and thus still represents a line ``separating"
the Higgs and confinement regions even in the absence of a phase
transition. At $\be_G=9$ we estimate it to be $\be_H^c=0.35406(5)$.

\begin{figure}
\begin{center}
\leavevmode
\epsfysize=250pt
\epsfbox[20 30 620 730]{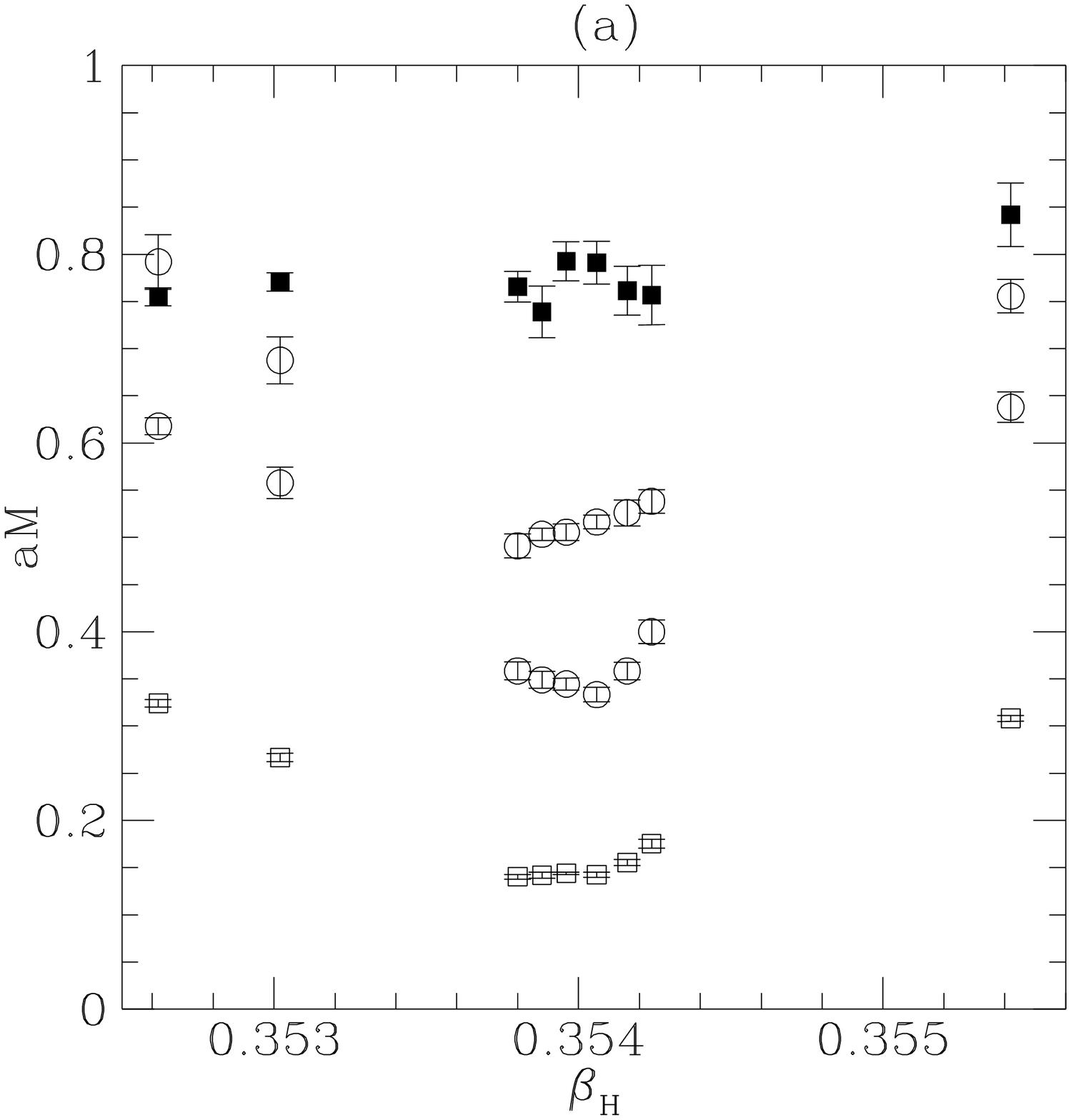}
\leavevmode
\epsfysize=250pt
\epsfbox[20 30 620 730]{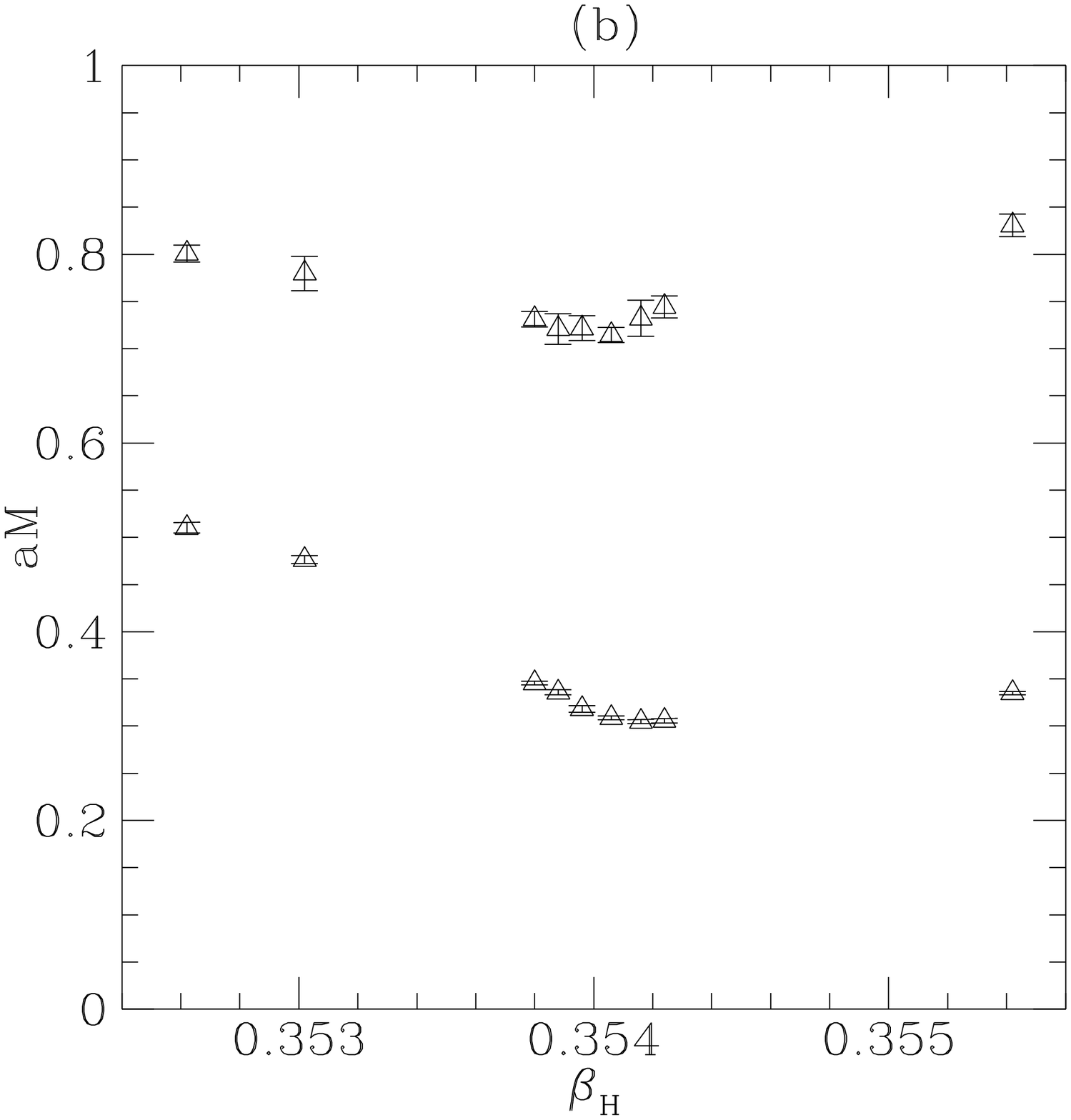}
\vspace{-1.6cm}
\end{center}
\caption[]{\it \label{mcross}
 Connecting the mass spectrum through the crossover.
 (a) The four lowest $0^{++}$ states. 
     Empty squares denote the scalar
     ground state and full squares the glueball. Circles correspond to 
     mixed intermediate states.
 (b) The two lowest $1^{--}$ states.}
\end{figure}

The variation of the lowest states in the $0^{++}$ and
$1^{--}$-channels along the crossover is shown in Fig.~\ref{mcross}.
The mass of the lightest state of the theory, the $0^{++}$ ground
state, shows a dip as the crossover is traversed. We have measured
this mass on lattices of size $L=24,36$ and found no lowering of the
mass with increasing lattice size. This means that essentially the
infinite volume mass is reached and that the physical ground state
mass definitely stays non-zero at the critical coupling. This
identifies the transition unambiguously as a
crossover\,\cite{bp94,kaj96}. It is interesting to note that the mass
of the third excited state, corresponding to the glueball on the
confinement side, does not change through the crossover region until
well into the Higgs region. This is yet another manifestation of its
decoupling, this time it proves stable under variation of another
scalar coupling parameter, the hopping parameter $\be_H$.

In the $1^{--}$ channel we show only the lightest state and 
the first excitation, as our basis is smaller in this channel.
The lowest state clearly becomes lighter as one changes from the confinement
region to the Higgs region.

\begin{figure}
\begin{center}
\leavevmode
\epsfysize=250pt
\epsfbox[20 30 620 730]{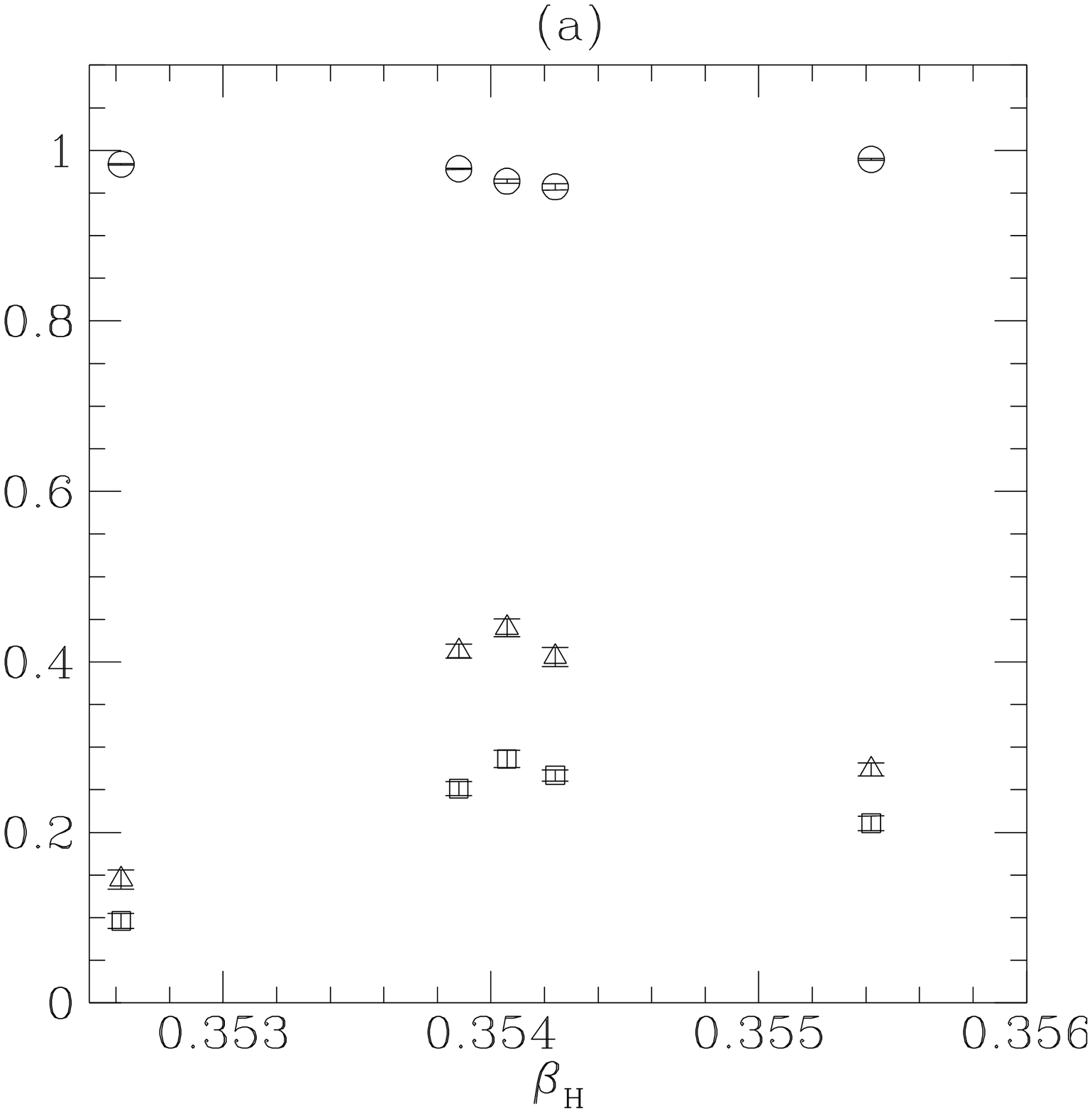}
\leavevmode
\epsfysize=250pt
\epsfbox[20 30 620 730]{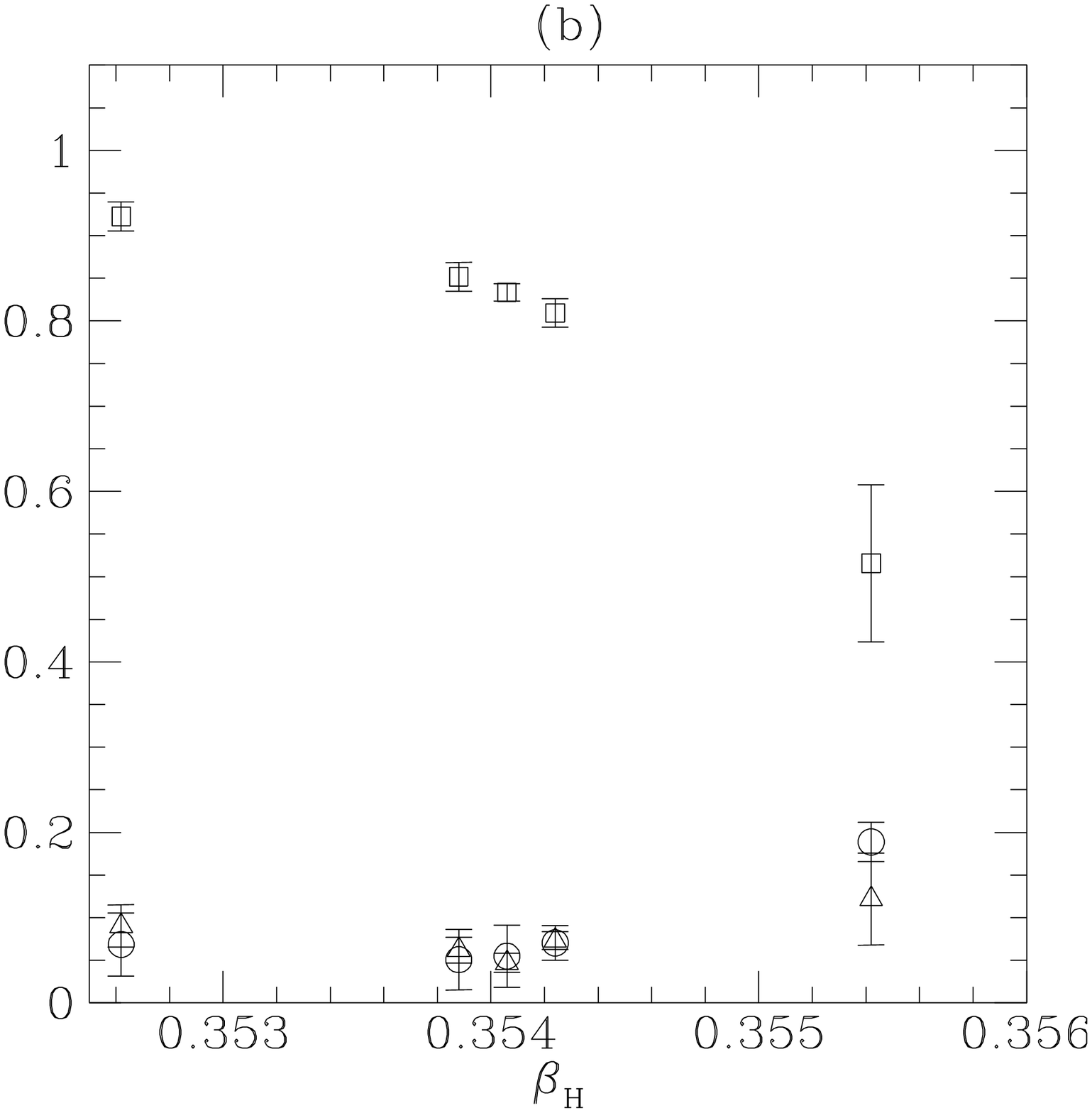}
\vspace{-1.6cm}
\end{center}
\caption[]{\it \label{crosscomp}
 Operator content through the crossover in the $0^{++}$-channel.
 a) The ground state,
 b) The glueball.
 Circles denote the maximal $a_{ik}$ from $R/L$ contributions, 
 squares those from $C$ and triangles represent the admixture of 
 $P^{(L)}$. }
\end{figure}

When the parameter $\be_H$ is varied through the crossover some states
change their ordering in mass, so that we felt unable to assign
distinct symbols to some of the states in Fig.\,\ref{mcross}\,(a). In
these cases the question arises how one identifies the individual
states. A useful criterion is to look at the composition of the mass
eigenstates in terms of the original operators used.  As an example,
we show the variation of the operator content of the $0^{++}$ ground
state and glueball through the crossover region in
Fig.~\ref{crosscomp}.
The figure exhibits a remarkable phenomenon concerning the mixing of
operators. As one moves from the confinement region towards the Higgs
region, the scalar ground state picks up more contributions from
purely ``gluonic'' operators. The mixing of scalar and gluonic
operators is strongest in the middle of the crossover, at the critical
coupling $\be^c_H$. This is not unexpected given the fact that here
the fluctuations are strongest. Although there is now a significant
gauge contribution to the lowest state, it still remains clearly
dominated by scalar operators. Moving into the Higgs region the gauge
contribution declines again and settles at some low value.  Looking at
the composition of the glueball, however, we do not see the analogous
feature happening. In contrast, the glueball maintains its almost pure
gauge composition throughout the crossover beyond the critical
coupling, with equally low overlap with the Polyakov loop. Only as one
moves clearly into the Higgs phase does the glueball give up its
purely gluonic nature, consistent with the decoupling observed from
the behaviour of its mass.
\begin{figure}
\begin{center}
\leavevmode
\epsfysize=25cm
\epsfbox[52 30 620 720]{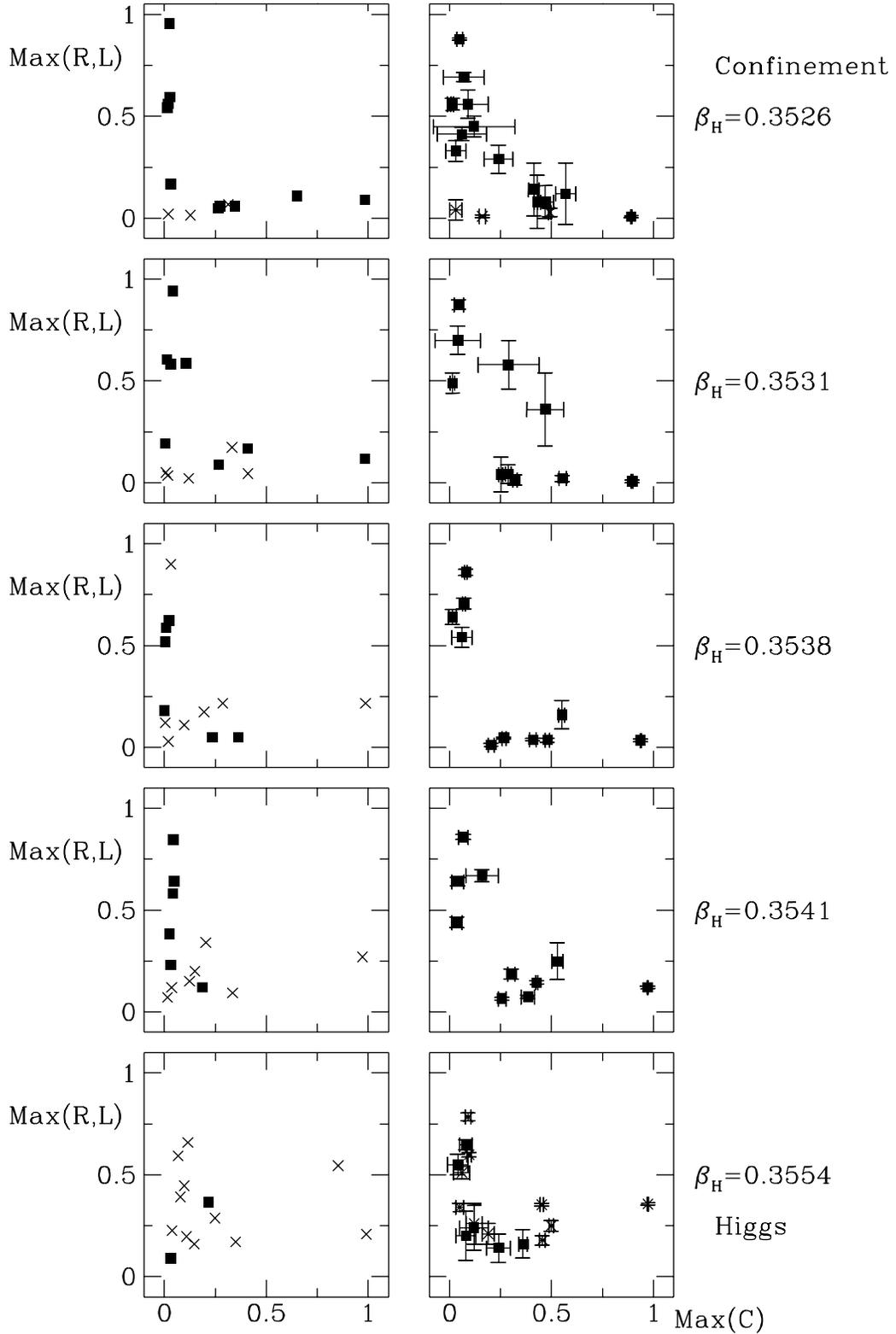}
\vspace{-4.8cm}
\end{center}
\caption[]{\it \label{scatter}
 Variation of the operator content of the $0^{++}$ (left) and $2^{++}$ (right)
 spectrum through the crossover as described in the text. 
 }
\end{figure}

Given the intrinsic interest of this decoupling phenomenon it is worth
investigating it in more detail. To do so we take the lowest 13 states
in the $0^{++}$ sector and display for each state the maximum overlap
onto scalar ($R$ or $L$) operators as well as the maximum overlap onto
the Wilson loop $(C)$ operators. We do so separately for our extreme
confining and Higgs coupling values as well as for three values of
$\beta_H$ that correspond to traversing the crossover region from near
the edge (on the confining side) to near the pseudo-critical value,
$\beta^c_H$. The corresponding plots are shown on the left of
Fig.~\ref{scatter}. (Errors are not shown since they are small.)
Those points shown as crosses correspond to states with substantial
overlap onto flux loop operators. If the states had random
scalar-gauge content, the points would be scattered over the whole
plot. On the other hand, if there was a total scalar-gauge decoupling,
then one or the other coordinate of each point would be zero. It is
clear from Fig.~\ref{scatter} that the latter is close to being the
case in the confining phase and that this persists, although more
weakly, as we traverse the crossover region. In the Higgs phase,
however, any decoupling effect is clearly much weaker. Note the
increasing number of states with a substantial flux loop overlap, as
we traverse the crossover towards the Higgs region. This is a
manifestation of the increasing instability of the confining flux tube
that we shall discuss in more detail below.

Also in Fig.~\ref{scatter}, on the right, we show similar plots for
the $2^{++}$. In this case we show the errors because several of them
are large. The reason for such large errors is that we have two almost
degenerate states whose order is inverted in some measurements,
relative to other measurements. That is to say, any point with large
errors is definitely not an approximate state of the system and should
hence be disregarded. If we do disregard those ``states" then we see
that we have a picture very similar to the one we obtained for the
$0^{++}$: a marked scalar-gauge decoupling in the confining region
that only begins to weaken at the very centre of the crossover
region. Indeed there seems to be some remnant of it even in the Higgs
phase.

\section{Flux-tube decay}

When comparing the spectrum on the Higgs and confinement side of the
first-order phase transition at small scalar coupling we discussed the
different interpretations that the mass governing the exponential
fall-off of the Polyakov line operator has: on the Higgs side it can
be identified with the volume-independent mass of a two-$W$ scattering
state while on the confinement side it rises linearly with the lattice
size and represents a flux loop. These different behaviours represent
the different dynamics in the two regimes. In this section we try to
answer the question how the physics changes from the former
interpretation to the latter. We know that the two regimes are
analytically connected, and so are the mass eigenstates of the Higgs
and confinement regimes. Hence, there must also be a smooth transition
in the projection of the Polyakov line operator.

First, we note that we lose the signal for the mass $aM_{P}$ as we
move into the crossover region from either side. At the end-points in
both regimes we do see plateaux in the effective mass, but they are
lost when moving closer towards $\be^c_H$. The situation is not
improved by increasing the statistics. Second, let us recall that the
Polyakov loop operator has an increasing projection onto the scalar
ground state towards the middle of the crossover. These two
observations may be given a physical interpretation. On the
confinement side of the crossover we saw already a small non-vanishing
expectation value for the Polyakov loop operator. Nevertheless, the
mass extracted from the exponential fall-off of its correlator still
exhibits the linear dependence on the lattice size and hence
represents a flux loop winding around the lattice, despite the fact
that our model permits screening, due to the presence of matter
fields. As we move to the critical point the Polyakov loop develops a
large expectation value which signals the loss of linear confinement,
and the fact that flux loops no longer exist. The question then is,
what happens to the flux loop as the parameters are changed? The
increasing overlap of the Polyakov loop with the $0^{++}$ ground state
suggests a natural explanation: As we move out of the confining region
the flux loop becomes more and more unstable, as it is increasingly
easy to pair-produce scalars that screen the flux. The $0^{++}$ ground
state is one possible decay product, just based on quantum numbers. A
stronger mixing between this state and the flux loop may thus indicate
the increasing instability of the latter.

\begin{figure}
\begin{center}
\leavevmode
\epsfysize=250pt
\epsfbox[20 30 620 730]{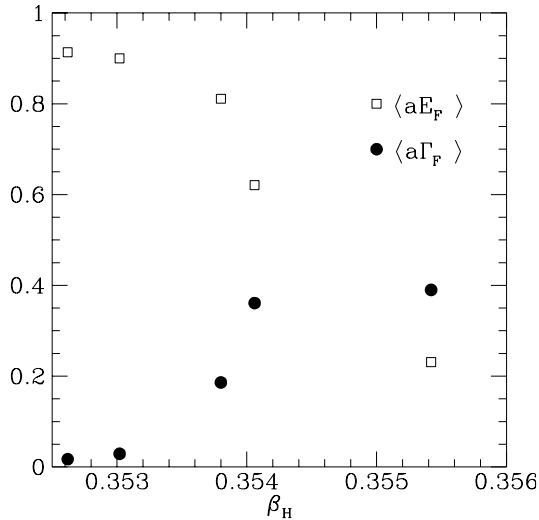}
\vspace{-1.6cm}
\end{center}
\caption[]{\it \label{gamma}
 Effective energy and decay width of the flux loop for the decay into
 single particle $0^{++}$ states. }
\end{figure}

In order to elaborate on this point we consider the overlap of the
Polyakov loop with all the $0^{++}$ eigenstates. If the suggested
picture is true then all these states are possible decay products. Of
course, there are other decay products as well, like pairs of $1^{--}$
particles etc.  For a first qualitative test of our picture it is
however sufficient to consider just the single particle states.  Let
us now treat the flux loop like a decaying resonance.  As long as it
is stable, its two-point correlation function has a pole at its
mass. When the resonance becomes unstable the pole is shifted away
from the real axis. The real part of the pole is then equal to the sum
of the masses of the decay products, weighted by the branching ratios
for the particular decay channel.  In analogy to this consideration we
now define an ``effective flux loop mass" by the sum of the masses of
the $0^{++}$ eigenstates, weighted by their maximal overlap onto a
Polyakov line operator, 
\beq
 \langle aE_F \rangle \equiv \frac{\sum_i
\mid a_{PL,i} \mid ^2 aM_i} {\sum_i \mid a_{PL,i} \mid ^2} \;. 
\eeq
In the same manner we construct the average square of the flux loop
energy, $\langle E_F^2 \rangle$. From these we then obtain the
``width" of the flux loop as the root mean square fluctuation 
\beq
(a\Gamma_F)^2=\langle (aE_F)^2 \rangle - \langle aE_F \rangle ^2\;.
\eeq 
These two quantities are plotted through the crossover region in
Fig.~\ref{gamma}.
Indeed, the decay width defined in this way is nearly zero on the
confinement side and then increases smoothly as it approaches the
critical coupling. It then maintains its higher value on the Higgs
side indicating the instability and non-existence of flux loops on
this side of the transition.

We conclude that, although the spectra are very similar on both sides
of the critical $\be^c_H$ for large scalar coupling, we have clear
evidence for a pronounced change in the dynamics as one moves from the
confining to the Higgs region. In particular it is possible to
demonstrate the growing instability of flux loops through the
transition. We remark, however, that the concept of a flux loop only
exists in a finite volume, and thus this analysis does not have an
infinite-volume limit. Nevertheless we would expect the same phenomena
to happen if one were to investigate the physical flux tube between
two static external charges at a given separation and then change the
parameters from the confining regime to the Higgs regime.

\section{Summary and conclusions}

In this paper we have extended our previous work on the
$(2+1)$-dimensional SU(2) gauge-fundamental Higgs model in various
ways.  Firstly, we have added the $2^{++}$ mass spectrum to our
previous calculations of the $0^{++}$ and $1^{--}$ spectra. Secondly
we have added operators based on Polyakov loops. If we have
confinement then these loops will project onto chromoelectric flux
loops (``torelons'') that close through one of the periodic spatial
boundaries.  In that case the lightest mass extracted from the
correlation function of such an operator will be proportional to the
lattice spatial size, and its mass density will give the confining
string tension (up to corrections, the leading one of which is
universal and known). This allows us to probe directly the confining
properties of the theory at various points in the phase diagram. At
the same time we have extended our basis of operators in the mass
spectrum calculations; in part by the inclusion of these torelon
operators and also products of them. This provides us with a much
better control of finite-volume effects. We have in addition performed
the calculations not only for the small scalar self-coupling at which
our previous simulations were carried out, but also for a large value
of the scalar coupling where no phase transition separates the
symmetric and Higgs phases. Here we have performed calculations
through the crossover region that separates these two phases.  In all
these calculations we have used only a single value of the inverse
gauge coupling, but one for which our earlier calculations assure us
that lattice corrections to the continuum limit are already small.

As well as extending our previous calculations, we have also
shifted the emphasis of our physics interest away from 
the Electroweak Theory at finite temperature to dynamical 
phenomena that may be relevant to that other theory
with gauge fields and charges in the fundamental 
representation: Quantum Chromodynamics. 

So one major focus of our interest has been confinement. At small
scalar self-coupling we find that the torelon mass increases
(approximately) linearly with its length and hence that the symmetric
phase is confining.  Of course we know that, just as in QCD, flux
tubes will be broken through the pair creation of fundamental charges,
but within our errors we see no sign of this. The string tension is
almost identical to its pure gauge value. By contrast, in the Higgs
phase, we see what perturbation theory tells us to expect: the
Polyakov loop operator couples to a $W\,W$-scattering state instead of
to the now non-existent flux tube. Indeed, apart from the Higgs
scalar, $W\,W$-scattering states are all we see in the $0^{++}$ and
$2^{++}$ spectra. More intriguing is the situation for large scalar
self-coupling. Here we once again have a symmetric phase that shows
linear confinement and a Higgs phase where linear confinement is
absent. However, now there is no first-order transition between the
two phases, only a smooth crossover. This raises the question of how
the dynamics can smoothly interpolate between the two phases. We
provided evidence for the following picture. In the pure gauge theory
a flux loop joined upon itself through a boundary is a stable state
(of the finite volume Hamiltonian). With fundamental charges present
such a state can decay.  Far from the crossover such decays are
strongly suppressed and the flux loop will be almost stable: a
resonance with a very narrow decay width. As we enter the crossover
region the decay width becomes rapidly larger: so the pole in the
complex energy plane moves rapidly away from the real axis. As
(initially) a secondary effect, the real part will also change,
because of the newly enhanced intermediate states, and so the mass
will shift as well. At some point, near the centre of the crossover,
the pole is so far from the real axis that one completely loses a
particle interpretation.  At this point it is no longer useful to talk
of confining flux tubes.  We have found it convenient to use a
periodic flux loop on a finite volume as a probe of confinement, but
one can clearly use a finite flux line between fundamental
sources/charges equally well.

Our second major interest has been the approximate decoupling between
scalar and gauge degrees of freedom which we observed in our previous
work. There we found that in the symmetric phase there are states
which are almost entirely gluonic and the masses of these states
coincide with the glueball masses of the pure gauge theory. We now
find that this occurs not only in the $0^{++}$ sector, but also in the
$2^{++}$ sector, and at large scalar self-coupling as well as at
small. Moreover this phenomenon only gradually disappears as we move
through the crossover to the Higgs phase. Indeed, there is some
evidence that a remnant of this phenomenon survives even in the Higgs
phase. Amongst other things this tells us that the mixing between
glueballs and states composed of fundamental charges is strongly
suppressed, at least as long as we have some remnant of linear
confinement. It also tells us that this is a very robust phenomenon.
Is it robust enough to extend to QCD? There it would support the idea
that glueball-quarkonium mixing should be suppressed and that we
should search experimentally for glueballs at the masses found in the
corresponding pure gauge theory.  These are important questions but we
need some understanding of the underlying dynamics if we are to move
beyond the realm of conjecture.

In addition to the above two topics, we also clarified the unusually
strong finite-volume effects that we saw in our previous work. It
turns out that these are operator artefacts. In fact, the true bound
states show weak finite volume effects. However, if one uses smeared
operators it is difficult to avoid components that are essentially
(smeared) Polyakov loops. These components introduce the torelon into
the measured spectrum and, of course, the mass of this varies rapidly
with volume.  This is a generic problem when one has a theory with
fundamental charges, and the simplest way to deal with it is to
include smeared torelon operators in the operator basis. After
diagonalisation one can explicitly determine which states have
dominant torelon components.

Finally we point to the dense spectrum of states we have found in the
Higgs phase at large scalar coupling. This is in a phase without
confinement, and presumably the dynamics is not being driven by the
gauge coupling becoming large at large distances. It is therefore
tempting to conclude that what we have here is a generic example of a
spectrum produced by a strong scalar self-coupling.

We conclude with a brief comment as to how these calculations need to
be improved. The most obvious lacuna concerns multiparticle scattering
states. We have not attempted to identify these, except in the very
obvious case of the Higgs phase at small scalar self-coupling.  There
is some evidence that our operators, which are naively designed to
project only onto single particle states, usually do so. However, this
really needs to be checked explicitly, especially in the
strong-coupling Higgs phase where, if it turns out not to be the case,
it could completely alter our interpretation of the spectrum. The
inclusion of scattering states is also necessary if we are to perform
a proper analysis of flux tube decay through the crossover region.
Technically the problem is simple. For example, suppose
$\phi(\vec{p},t)$ is an operator that has a very good projection onto
the lightest scalar with momentum $\vec{p}$ (which will be one of the
discrete set of allowed momenta in our finite volume). Then the
operator $\phi_2(t) \equiv \phi(+\vec{p},t) \phi(-\vec{p},t)$ (vacuum
subtracted if necessary) will be a good operator for the two-scalar
scattering state of zero total momentum and finite relative
momentum. At least this will be so for volumes that are not too
small. In similar ways one can build other trial multiparticle
operators and add them to the operator basis. However, to be confident
that one is correctly isolating multiparticle states one needs to do
an accurate finite volume analysis that shows the continuum cut
gradually forming as one increases the volume. This would be a
substantially larger calculation than the one in this paper. It would
also be very interesting to further pursue the scalar-glue decoupling;
in particular to see if it might be related to the OZI rule in QCD.

\paragraph{Acknowledgements}

These calculations were performed on the Cray J90 at RAL under PPARC
grants GR/J86605 and GR/K9533. We thank Chris Sachrajda for allowing
us to use his computer time allocation on this machine. We also
acknowledge support under PPARC grant GR/K55752.


\newpage
\begin{thebibliography}{99}

\bibitem{kaj95}
K.~Kajantie, M.~Laine, K.~Rummukainen and M.~Shaposhnikov,
Nucl.~Phys.~B466 (1996) 189.
%
\bibitem{kaj96}
K.~Kajantie, M.~Laine, K.~Rummukainen and M.~Shaposhnikov,
Phys.~Rev.~Lett.~77 (1996) 2887.
%
\bibitem{ilg95}
E.M.~Ilgenfritz, J.~Kripfganz, H.~Perlt and A.~Schiller,
Phys.~Lett.~B356 (1995) 561;\\
M.~G\"urtler, E.M.~Ilgenfritz, J.~Kripfganz, H.~Perlt and A.~Schiller,
Nucl.~Phys.~B83 (1997) 383;\\
M.~G\"urtler, E.M.~Ilgenfritz and A.~Schiller, Eur.~Phys.~J.~C1 (1998)
363; Phys.~Rev.~D56 (1997) 3888.
%
\bibitem{us}
O.~Philipsen, M.~Teper and H.~Wittig, Nucl.~Phys.~B469 (1996) 445.
%
\bibitem{Bielefeld} 
F.~Karsch, T.~Neuhaus and A.~Patk\'os, Nucl.~Phys.~B442 (1995) 629;\\
F.~Karsch, T.~Neuhaus, A.~Patk\'os and J.~Rank, Nucl.~Phys.~B474
(1996) 217.

\bibitem{fkrs94}
P. Ginsparg,
Nucl.\ Phys.\ B 170 (1980) 388;
T. Appelquist and R. Pisarski,
Phys.\ Rev.\ D 23 (1981) 2305;
K.\ Kajantie, M.\ Laine, K.\ Rummukainen and M.\ Shaposhnikov,
Nucl.\ Phys.\ B 458 (1996) 90;
A.~Jakov\'ac and A.~Patk\'os,
Nucl.\ Phys.\ B 494 (1997) 54.

\bibitem{mont97}
F.~Csikor, Z.~Fodor, J.~Hein, J.~Heitger, A.~Jaster and I.~Montvay,
Nucl.~Phys.~B (Proc. Suppl.) 53 (1997) 612.
 
\bibitem{bp94}
W. Buchm\"uller and O. Philipsen, Nucl.~Phys.~B443 (1995) 47.

\bibitem{frad79}
E.~Fradkin and S.~Shenker, Phys.~Rev.~D19 (1979) 3682.

\bibitem{rum96}
K.~Rummukainen, Nucl.~Phys.~B (Proc. Suppl.) 53 (1997) 30. 

\bibitem{do95}
H.-G.~Dosch, J.~Kripfganz, A.~Laser and M.~G.~Schmidt,
Phys.~Lett.~B365 (1996) 213; Nucl.~Phys.~B507 (1997) 519.
 
\bibitem{bp97}
W.~Buchm\"uller and O.~Philipsen, Phys.~Lett.~B397 (1997) 112.

\bibitem{lat96} 
O.~Philipsen, M.~Teper and H.~Wittig, Nucl.~Phys.~B
(Proc. Suppl.) 53 (1997) 626.

\bibitem{owe_eger_97}
O.~Philipsen, presented at E\"otv\"os Conference in Science: Strong and
Electroweak Matter (SEWM 97), Eger, Hungary, 21-25 May 1997, hep-ph/9708309.

\bibitem{GuIlSchiStr97} M. G\"urtler, E.M. Ilgenfritz, A. Schiller and
C. Strecha, Nucl.~Phys.~B (Proc. Suppl.) 63A--C (1998) 563.


\bibitem{lai95}
M.~Laine, Nucl.~Phys.~B451 (1995) 484.

\bibitem{tep87}
M.~Teper, Phys.~Lett.~B187 (1987) 345.
%
\bibitem{albanese}
M.\,Albanese et al., Phys.~Lett. B192 (1987) 163; Phys.~Lett. B197
(1987) 400.
%
\bibitem{how89}
H.\,Wittig, Nucl.~Phys. B325 (1989) 242.

\bibitem{cm87a}
C.\,Michael, J.~Phys. G13 (1987) 1001.

\bibitem{fabhaan}
K.\,Fabricius and O.\,Haan, Phys.~Lett. B143 (1984) 459.
 
\bibitem{kenpen}
A.D.\,Kennedy and B.J.\,Pendleton, Phys.~Lett. B156 (1985) 393.

\bibitem{bunk_lat94}
B.\,Bunk, Nucl.~Phys.~B (Proc. Suppl.) 42 (1995) 566.

\bibitem{MontWeisz87}
I.\,Montvay and P.\,Weisz, Nucl.~Phys. B290 (1987) 327.

\bibitem{for85}
P. de Forcrand G. Schierholz, H. Schneider and M. Teper,
Phys.~Lett.~160B (1985) 137.

\bibitem{tepunp}
M.~Teper, unpublished.

\bibitem{tep93}
M.~Teper, Phys.~Lett.~B311 (1993) 223.



\bibitem{bfh}
W.~Buchm\"uller and Z.~Fodor, Phys.~Lett.~B331 (1994) 131. 

\end{thebibliography}
\end{document}